\documentclass[twocolumn, twocolappendix]{aastex631}
\usepackage{amsmath}
\usepackage{multirow}
\usepackage{enumitem}
\usepackage{color}
\usepackage{ulem}
\usepackage{CJKfntef,CJK}

\graphicspath{{figures/}}

\newcommand\hcop{$\rm HCO^+$}
\newcommand\nhp{$\rm N_2H^+$}
\newcommand\coa{$\rm ^{12}CO$}
\newcommand\cob{$\rm ^{13}CO$}
\newcommand\coc{$\rm C^{18}O$}
\newcommand\hcn{$\rm HCN$}
\newcommand\hnc{$\rm HNC$}
\newcommand\kms{$\rm km\ s^{-1}$}
\newcommand\NhcoponNco{$N({\rm HCO^+})/N({\rm CO})$}
\newcommand\NhcoponNnhp{$N({\rm HCO^+})/N({\rm N_2H^+})$}
\newcommand\cobratio{$I(^{13}{\rm CO}\ J=2\text{--}1)/I(^{13}{\rm CO}\ J=1\text{--}0)$}

\begin{document}
\begin{CJK*}{UTF8}{gbsn}

\title{Shock and Cosmic Ray Chemistry Associated with the Supernova Remnant W28}

\author[0000-0002-9776-5610]{Tian-yu Tu (涂天宇)}
\affiliation{School of Astronomy \& Space Science, Nanjing University, 163 Xianlin Avenue, Nanjing 210023, China}

\author[0000-0002-4753-2798]{Yang Chen (陈阳)}
\affiliation{School of Astronomy \& Space Science, Nanjing University, 163 Xianlin Avenue, Nanjing 210023, China}
\affiliation{Key Laboratory of Modern Astronomy and Astrophysics, Nanjing University, Ministry of Education, Nanjing 210023, China}
\email{ygchen@nju.edu.cn}

\author[0000-0002-5683-822X]{Ping Zhou (周平)}
\affiliation{School of Astronomy \& Space Science, Nanjing University, 163 Xianlin Avenue, Nanjing 210023, China}
\affiliation{Key Laboratory of Modern Astronomy and Astrophysics, Nanjing University, Ministry of Education, Nanjing 210023, China}

\author[0000-0001-6189-7665]{Samar Safi-Harb}
\affiliation{Department of Physics and Astronomy, University of Manitoba, Winnipeg, MB R3T 2N2, Canada}



\begin{abstract}

Supernova remnants (SNRs) exert strong influence on the physics and chemistry of the nearby molecular clouds (MCs) through shock waves and the cosmic rays (CRs) they accelerate. 
To investigate the SNR-cloud interaction in the prototype interacting SNR W28 (G6.4$-$0.1), we present new observations of \hcop, \hcn\ and \hnc\ $J=1\text{--}0$ lines, supplemented by archival data of CO isotopes, \nhp\ and $\rm H^{13}CO^+$. 
We compare the spatial distribution and spectral line profiles of different molecular species. 
Using local thermodynatic equilibrium (LTE) assumption, we obtain an abundance ratio \NhcoponNco\ $\sim10^{-4}$ in the northeastern shocked cloud, which is higher by an order of magnitude than the values in unshocked clouds. 
This can be accounted for by the chemistry jointly induced by shock and CRs, with the physical parameters previously obtained from observations: preshock density $n_{\rm H}\sim 2\times 10^{5}\rm \ cm^{-3}$, CR ionization rate $\zeta=2.5\times 10^{-15} \rm \ s^{-1}$ and shock velocity $V_{\rm s}=15\text{--}20\rm \ km\ s^{-1}$. 
Towards a point outside the northeastern boundary of W28 with known high CR ionization rate, we estimate the abundance ratio $ N({\rm HCO^+})/N({\rm N_2H^+}) \approx 0.6\text{--}3.3$, which can be reproduced by a chemical simulation if a high density $n_{\rm H}\sim 2\times 10^5 \ \rm cm^{-3}$ is adopted.

\end{abstract}

\keywords{Molecular clouds (1072) --- Supernova remnants (1667) --- Cosmic rays (329) --- Shocks (2086) --- Abundance ratios (11)}


\section{Introduction} \label{sec:intro}
When massive stars end their lives as energetic supernova (SN) explosions, many of them have not yet moved far away from their natal molecular clouds (MCs). 
The physical and chemical conditions in the MCs could be strongly modified by the energy feedback of the supernova explosions and the following supernova remnant (SNR) evolution. 
Many SNRs exhibit evidence of interaction with nearby MCs \citep[e.g.][]{Seta_Enhanced_1998,Jiang_Cavity_2010,Kilpatrick_Systematic_2016,Zhou_Systematic_2023}. 
These interactions are crucial to understanding the feedback of SNe and SNRs into the ISM. 

\par

The shock waves of SNRs propagating into dense MCs will compress and heat the gas \citep{Draine_Theory_1993}. 
The chemistry in MCs is sensitive to temperature and density, and thus sensitive to the passage of the shock wave. 
The shock chemistry has been extensively studied in shocked regions of protostellar outflow \citep[e.g.][]{Podio_Molecular_2014, Lefloch_L1157B1_2017}. 
There are also a few studies concerning the chemistry of SNRs. 
For example, \citet{vanDishoeck_Submillimeter_1993} presented a comprehensive discussion of the chemistry in the SNR IC443. 
\citet{Lazendic_Multiwavelength_2010a} found an enrichment of \hcop\ and SO in the SNR G349.7+0.2. 
\citet{Zhou_Unusually_2022b} found high abundance ratio between \hcop\ and CO in the SNR W49B and attributed it to shock wave, cosmic ray (CR) ionization, and possibly the X-ray precursor of the shock. 
\citet{Mazumdar_Submillimeter_2022a} revealed formation of $\rm H_2CO$ and $\rm CH_3OH$ in the shocked region of the SNR W28. 
However, observational study of shock chemistry is still scarce and more investigations are required.

\par

Meanwhile, SNR shocks are believed to be the prime accelerators of Galactic CRs following the diffuse shock acceleration process \citep{Bell_acceleration_1978}. 
Many SNRs are found to be located close to $\gamma$-ray sources \citep[e.g.,][and references therein]{Aharonian_Gamma_2013}. 
The high-energy emissions originate from either the decay of $\pi^0$ mesons produced by collision between high-energy ($\gtrsim 280\ \rm MeV$) CR protons with H nuclei in MCs, which is the so-called hadronic scenario, or the inverse Compton scattering of background radiation by high-energy electrons, which is the so-called leptonic scenario. 
In the case that the SNR is associated with MCs, the hadronic scenario can naturally explain the $\gamma$-ray emission if the observed $\gamma$-ray spectra can be reasonably fitted \citep[e.g.][]{Aharonian_Discovery_2008,Fukui_Pursuing_2021}.

\par

Low-energy CR protons will not contribute to $\gamma$-ray emission, but they are the dominating ionization sources in dense MCs into which ionizing UV emission is difficult to penetrate \citep{Padovani_Cosmic-ray_2009}. 
CR protons ionize molecular hydrogen gas mainly through:
\begin{equation}
    \rm H_2 \xrightarrow[]{CR\ protons} H_2^+ + e^-
\end{equation}
followed by
\begin{equation}
    \rm H_2^+ +H_2 \longrightarrow H_3^+ + H.
\end{equation}
The $\rm H_3^+$ ion starts various chemical reactions in MCs. 
Chemical simulations have shown that CR ionization exerts strong influence on the chemistry of MCs \citep[e.g.][]{Albertsson_Atlas_2018}. 
The usability of chemical codes and networks requires observational tests. 
However, few observations have been done concerning the chemical effects of high CR ionization rate, especially in the MCs associated with SNRs.

\par

SNR W28 is one of the prototypes for studying SNR-cloud interaction, with an estimated age of $\sim 3.3\text{--}4.2\times 10^4$ yr \citep{Velazquez_Investigation_2002,Rho_ROSAT_2002,Li_g-rays_2010} at a distance of $\sim 1.9$ kpc \citep{Velazquez_Investigation_2002,Ranasinghe_Distances_2022a}. 
Located in a complex of MCs, W28 is believed to be interacting with MCs. 
Evidence for the interaction includes molecular line broadening spatially coincidence with the radio continuum and TeV $\gamma$-ray emission \citep{Wootten_dense_1981, Arikawa_Shocked_1999,Reach_Shocked_2005,Nicholas_12_2011,Nicholas_mm_2012,Gusdorf_Probing_2012,Maxted_Ammonia_2016,Mazumdar_Submillimeter_2022a}, 1720 MHz OH masers \citep{Claussen_Polarization_1997, Hewitt_Survey_2008a}, infrared $\rm H_2$ emissions \citep{Reach_Infrared_2000, Reach_Shocked_2005, Yuan_Spitzer_2011}, etc. 
In the $\gamma$-ray band, GeV and TeV emissions are found associated with W28 \citep{Aharonian_Discovery_2008, Cui_Leaked_2018}, which are expected to originate from the hadronic scenario. 
The overall TeV emission shows two parts: HESS J1801$-$233 towards the northeastern MCs, and HESS J1800$-$240 towards the southern MCs outside the boundary of W28. 
The latter is further divided into three components: A, B and C, all of which are coincident with dense MCs. 
The $\gamma$-ray emission outside the radio boundary is ascribed to the illumination by the escaped CRs from W28 \citep[e.g.,][]{Li_g-rays_2010,Cui_Leaked_2018}. 

\par

\citet{Vaupre_Cosmic_2014} estimated the CR ionization rate with the abundance ratio $N({\rm DCO^+})/N({\rm HCO+})$ in some regions to be $\gtrsim 10^{15}\ \rm s^{-1}$, which is higher than the standard value by two orders of magnitude \citep{Glassgold_Model_1974a}. 
This makes W28 the third SNR directly measured to exhibit high CR ionization rate after IC443 \citep{Indriolo_Investigating_2010a} and W51C \citep{Ceccarelli_Supernova-enhanced_2011}.
Besides, the existence of 1720 MHz OH masers and class I methanol $\rm CH_3OH$ masers \citet{Pihlstrom_Detection_2014} also requires high CR ionization rate \citep{Nesterenok_Modelling_2022}. 
In addition, discovery of the Fe I K$\alpha$ line at 6.4 keV \citep{Nobukawa_Evidence_2018,Okon_origin_2018} further indicates the ionization induced by low energy CRs. 
All of the mentioned observations render W28 an ideal site to study the chemistry of SNR shock and CR ionization.  

\par

The $J=1$--0 lines of \hcop, \hcn\ and \hnc\ molecules are typical tracers of dense gas \citep[e.g.][]{Shirley_Critical_2015}. 
The abundance ratio between \hcop\ and other specific molecules, for example, CO \citep[e.g.,][]{Zhou_Molecular_2018,Bisbas_PDFCHEM_2023}, has been proposed to be a tracer of CR ionization \citep[e.g.][]{Albertsson_Atlas_2018}, while the line ratio $I({\rm HCN})/I({\rm HNC})$ serves as a tracer of gas kinetic temperature \citep[e.g.][]{Hacar_HCNtoHNC_2020}. 
Therefore, observation of these three lines towards W28 provides information of the physical and chemical effects of W28 on its adjacent MCs. 

\par

In this paper, we present new \hcop, \hcn\ and \hnc\ $J=1\text{--}0$ observations towards W28. 
Using the new data together with the archival data and chemical models, we investigate the chemistry brought by shock wave and CRs. 
The paper is organized as follows. 
In Section \ref{sec:obs}, we describe the new observations and archival data. 
In Section \ref{sec:res}, we present the results of the observations. 
We derive the abundance ratios between molecular species, and then present the results and relevant discussions of chemical simulations in Section \ref{sec:disc}. 
The conclusions are summarized in Section \ref{sec:con}.

\section{Observations} \label{sec:obs}
\subsection{\hcop, \hcn\ and \hnc\ observations and data reduction}

\begin{figure*}[htbp]
\plotone{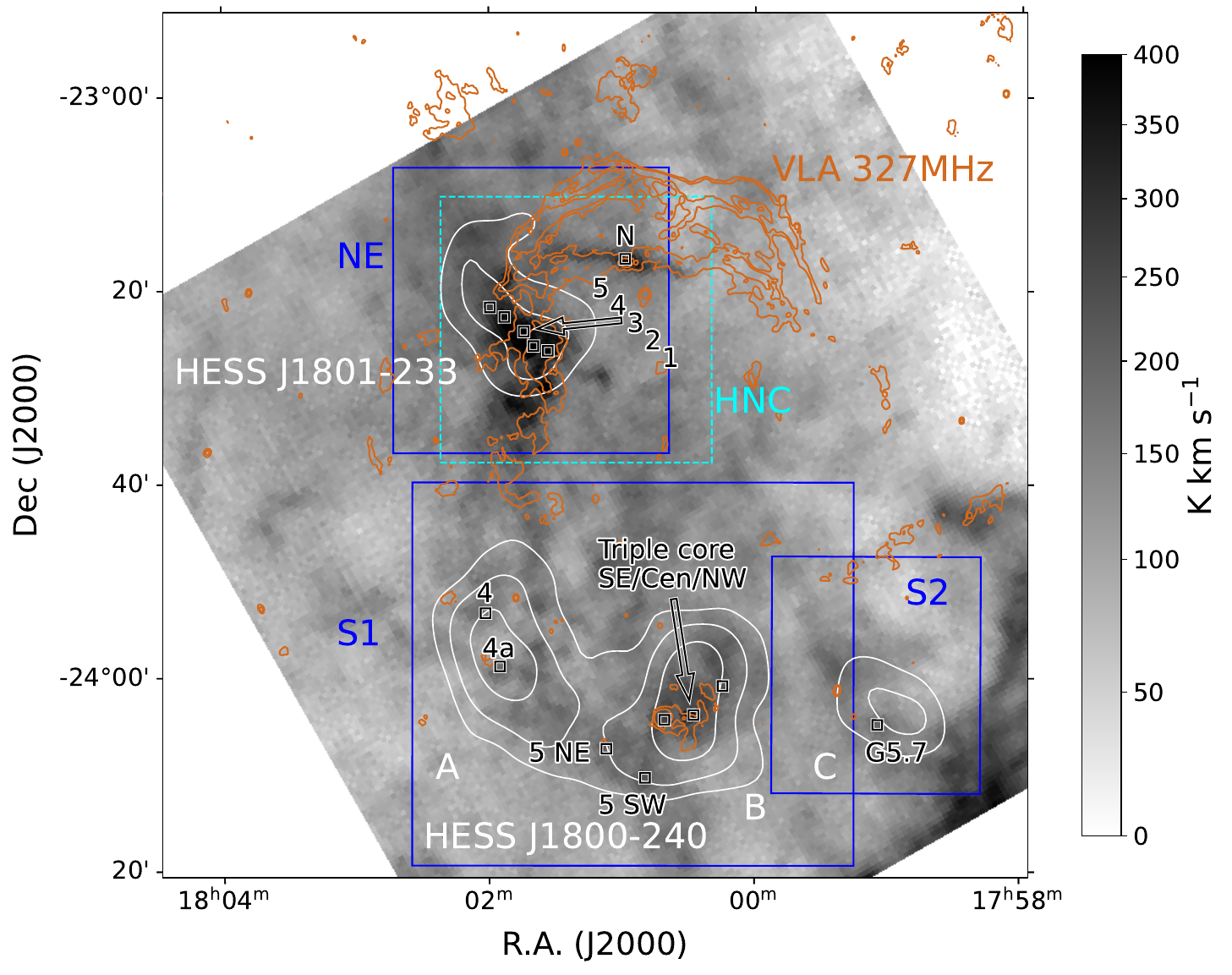}
\caption{Integrated intensity of \coa\ $J=1\text{--}0$ from $-50$ \kms\ to $+70$ \kms. The brown contours show the VLA 327 MHz continuum emission (levels are 60 and 180 $\rm mJy\ beam^{-1}$).  White contours trace the TeV emission as seen by HESS (levels are 4--6$\sigma$), with HESS J1801$-$233 in the northeast and HESS J1800$-$240 in the south (A, B and C of HESS J1800$-$240 are arranged from east to west). The blue boxes delineate the regions where we observed with PMOD the \hcop\ and \hcn\ $J=1\text{--}0$ lines, while the dashed cyan box delineates the region where we observed the \hnc\ $J=1\text{--}0$ line. The black boxes mark the regions from which we extract the spectra (see Figures \ref{fig:w28_ne_spec} and \ref{fig:w28_s_spec})
\label{fig:w28all}}
\end{figure*}

Observations of the $J=1\text{--}0 $ emission lines of \hcop, \hcn\ and \hnc\ were performed with the 13.7m millimeter-wavelength telescope of the Purple Mountain Observatory at Delingha (hereafter PMOD; PI: Yang Chen). 
The \hcop\ and \hcn\ lines were simultaneously observed with the SIS receiver in two epochs during 2018 June and 2021 August--October. 
These observations used on-the-fly (OTF) mapping to cover the northeastern MCs of W28 with a mapping area of $28^\prime \times 29^\prime$ and the southern MCs of W28 with two mappings, each with an area of $40^\prime \times 45^\prime$ and $21^\prime \times 24^\prime$. 
\hnc\ observations were conducted separately in 2021 June--August to cover the northeastern MCs of W28 with a mapping area of $28^\prime \times 27^\prime$. 
Positions and coverage of the regions are shown in Figure \ref{fig:w28all}.
The Fast Fourier Transform Spectrometers with 1 GHz bandwidth and 16,384 channels were used as the back ends, providing a velocity resolution of 0.21 \kms\ at 89 GHz. 
The half-power beamwidth (HPBW) of the telescope at 89 GHz is $\approx 60^{\prime\prime}$. 
The main beam efficiency was around 0.60 in 2018 and 0.58 in 2021\footnote{\url{http://www.radioast.nsdc.cn/zhuangtaibaogao.php}}. 
The typical system temperature $T_{\rm sys}$ is between 150 and 310 K, depending on the weather condition and the elevation of the source. 
The pointing accuracy of the antenna better than $5^{\prime\prime}$. 
The raw data were reduced with the GILDAS/CLASS package\footnote{\url{https://www.iram.fr/IRAMFR/GILDAS/}}. 
The data cubes of \hcop, \hcn\ and \hnc\ were all resampled to have the same velocity channel width of 0.25 \kms\ and the same pixel size of $30^{\prime\prime}$. 
The RMS noise is $\sim0.1$ K for \hcop\ and \hcn\ and $\sim 0.05$ K for \hnc.

\subsection{Other archival data}
We used other archival data to support our analysis for a multiwavelength view. 
We obtained \coa, \cob\ and \coc\ $J=1\text{--}0$ data from the Milky Way Image Scroll Painting (MWISP) line survey project. 
The HPBW is about $50^{\prime\prime}$ at 115 GHz and the pixel size is $30^{\prime\prime}$. 
The velocity channel width is 0.16 \kms\ for \coa\ and 0.17 \kms\ for \cob\ and \coc, and the typical noise measured in $T_{\rm mb}$ is 0.5 K for \coa\ and 0.3 K for \cob\ and \coc. 
Detailed description of the project can be found in \citet{Su_Milky_2019}. 
\par

We also retrieved the Very Large Array (VLA) 327 MHz continuum data from the MAGPIS (The Multi-Array Galactic Plane Imaging Survey) website\footnote{\url{https://third.ucllnl.org/gps/index.html}}. 
The \cob\ $J=2\text{--}1$ data cube was taken from the SEDIGISM \citep[Structure, Excitation and Dynamics of the Inner Galactic Interstellar Medium,][]{Schuller_SEDIGISM_2021} survey, which is observed with the APEX telescope with an angular resolution of $30^{\prime\prime}$ and a sensitivity of 0.8--1.0 K at 0.25 \kms\ velocity resolution. 
Supplementary data cubes of $\rm H^{13}CO^+$ and \nhp\ were taken from the MALT90 survey \citep[Millimetre Astronomy Legacy Team 90 GHz Survey,][]{Jackson_MALT90_2013a} with angular resolution of $\approx 38^{\prime\prime}$ and a pixel size of $\approx 9^{\prime\prime}$. 
The typical sensitivity measured in antenna temperature ($T_{\rm A}^*$) is 0.2--0.25 K at a velocity resolution of 0.11 \kms. 
The antenna temperature is transferred to main beam temperature $T_{\rm mb}$ with a main beam efficiency of 0.49.
The Herschel column density map was obtained from \citet{Marsh_Multitemperature_2017a} who fitted the data of Hi-GAL (Herschel infrared Galactic Plane) survey in 70--500 $\mu$m with the PPMAP procedure \citep[point process mapping,][]{Marsh_Temperature_2015}. 
The angular resolution of the map is $12^{\prime\prime}$. 

\par

All the processed data were further analyzed with \emph{Python} packages Astropy \citep{AstropyCollaboration_Astropy_2022} and Spectral-cube \citep{Ginsburg_Radio_2015}.
The data cubes were reprojected with Montage\footnote{\url{http://montage.ipac.caltech.edu/}} package when necessary. 
We visualized the data with \emph{Python} package Matplotlib\footnote{\url{https://matplotlib.org/}}.

\section{Results} \label{sec:res}
\subsection{Northeastern molecular clouds/HESS J1801$-$233}

\begin{figure}
\centering
\includegraphics[width=0.47\textwidth]{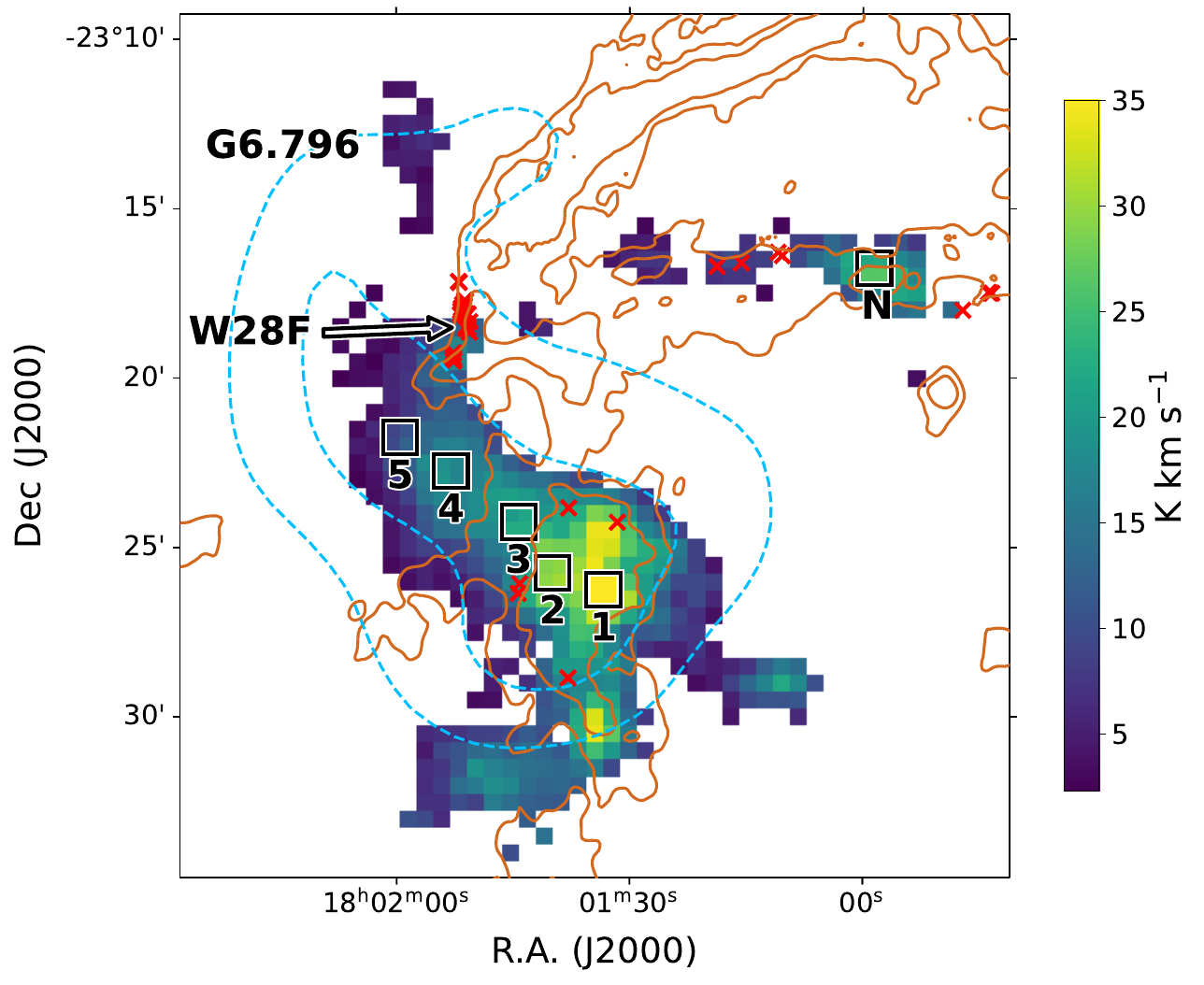}
\caption{Integrated intensity map of \hcop\ $J=1\text{--}0$ from $-30$ \kms\ to $+40$ \kms\ in the NE region (as delineated in Figure \ref{fig:w28all}). The brown contours show the VLA 327 MHz continuum emission (levels are 60 and 180 $\rm mJy\ beam^{-1}$). The dashed blue contours trace the TeV emission of HESS J1801$-$233 (levels are 4$\sigma$ and 5$\sigma$). The red crosses are 1720 MHz OH masers detected by \citet{Claussen_Polarization_1997}. The black boxes are regions where the \hcop\ spectra were extracted.
\label{fig:w28_ne_hcop}}
\end{figure}

In Figure \ref{fig:w28_ne_hcop}, we show the integrated intensity map of \hcop\ from $-30$ \kms\ to $+40$ \kms\ 
in the northeastern (NE) region overlaid with VLA 327 MHz continuum and HESS TeV $\gamma$-ray emission. 
The range of local-standard-of-rest (LSR) velocity is chosen according to the velocity span of the shocked \hcop\ emission. 
This velocity interval is consistent with the broadened line profile observed with other molecular transitions \citep{Reach_Shocked_2005,Nicholas_12_2011,Nicholas_mm_2012}. 
The broad line emission of \hcop\ is generally aligned with the shock front traced by radio continuum, which is solid evidence of shock-cloud interaction. 
Notably, the \hcop\ emission also appears essentially coincident with the TeV emission, which further supports the hadronic scenario of the $\gamma$-ray emission and provides a hint for CR ionization. 
The main body of the \hcop\ emission is in the center of the figure, while small portions of emission are also found in the northwestern (surrounding the region N) and northeastern (outside the radio emission) of the figure. 
The emission outside the radio continuum is spatially coincident with a massive star-forming region, G6.796$-$0.256 (hereafter G6.796 for short), in the ATLASGAL survey \citep{Urquhart_ATLASGAL_2018}, and is observed by the MALT90 survey.
We will discuss this region in Section \ref{G6.796-0.256}. 

\par

The main body of the \hcop\ emission shows bright clumps coincident with the brightest part of the radio continuum spatially covering Region 1. 
The integrated intensity decreases towards the northeast. 

\par

Although the 1720 MHz OH masers are generally within the spatial extent of the \hcop\ emission, most of them are not coincident with the brightest part of the \hcop\ emission. 
The \hcop\ emission found towards W28F, the northeastern part of the radio continuum where a cluster of OH masers is located, is not so bright as the emission around Region 1. 
However, observations of \citet{Gusdorf_Probing_2012} show bright high-J (up to $J_{\rm up}=11$) CO emission towards this region, indicating strong shock disturbance and high temperature. 
The weak \hcop\ emission may be attributed to the higher energy level population due to the high temperature, or beam dilution since the beam of PMOD is larger than that of \citet{Gusdorf_Probing_2012}.

\begin{figure*}
\centering
\includegraphics[width=0.98\textwidth]{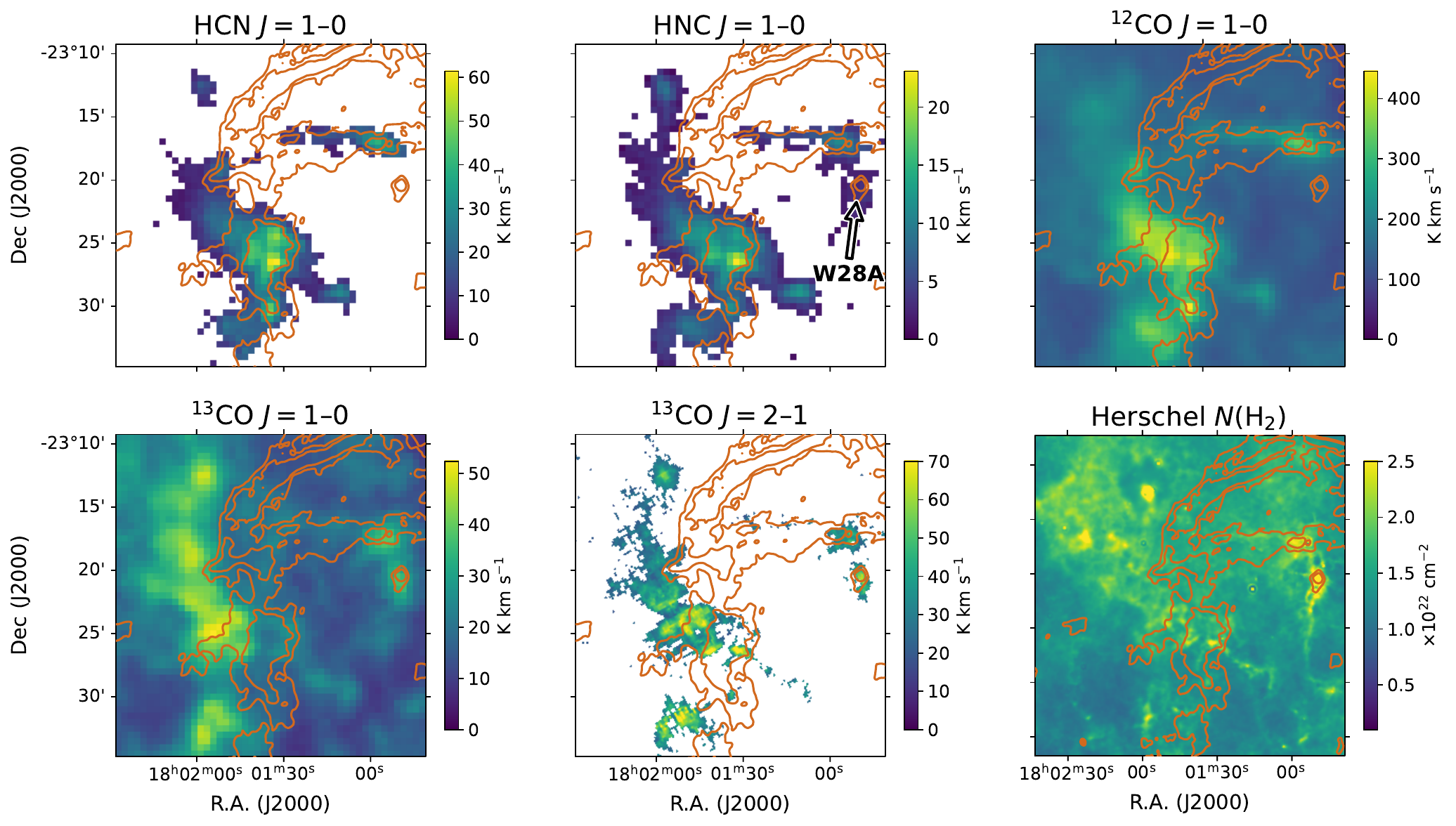}
\caption{Herschel column density map as well as integrated intensity of \hcn\ $J=1\text{--}0$, \hnc\ $J=1\text{--}0$, \coa\ $J=1\text{--}0$, \cob\ $J=1\text{--}0$, and \cob\ $J=2\text{--}1$ in the ``NE'' region. The velocity range of integration is $-30$ \kms\ to $+40$ \kms\ for \hcn, \hnc\ and \coa\, while $-15$ \kms\ to $+30$ \kms\ for \cob\ $J=1\text{--}0$ and $J=2\text{--}1$ lines. Pixels with integrated intensity weaker than $5\sigma$ are masked out. The light brown contours show the VLA 327 MHz continuum emission (levels are 60 and 180 $\rm mJy\ beam^{-1}$).
\label{fig:w28_ne_intensity}}
\end{figure*}

In Figure \ref{fig:w28_ne_intensity} we show the Herschel column density map as well as the integrated intensity maps of \hcn, \hnc, \coa\ and \cob\ $J=1\text{--}0$ and \cob\ $J=2\text{--}1$. 
Morphologies of the \hcn\ and \hnc\ emission are both similar to that of the \hcop\ emission. 
The \hnc\ emission is also detected in the westernmost part of the figure \citep[coincident with HII region W28A,][]{Goudis_W28_1976} and between the main body and G6.796. 
The non-detection of \hcop\ and \hcn\ in these regions might be attributed to lower sensitivity of the detector. 
The spatial distribution of \coa\ is also generally similar to that of \hcop, although \coa\ is much stronger and more extended. 
Emission of \cob\ $J=1\text{--}0$ is stronger outside the radio continuum than inside, which is different from the case of \hcop. 
The reason may be that \cob\ $J=1\text{--}0$ is often optically thin and traces the unshocked cloud \citep[e.g.][]{Su_Molecular_2011}. 
The \cob\ $J=2\text{--}1$ data has fine angular resolution and shows detailed structure of the molecular clouds. 
Several clumpy structures are seen towards the margin of the radio continuum. 
The distribution of \cob\ $J=2\text{--}1$ is broadly coincident with the Herschel column density map.

\begin{figure*}
\plotone{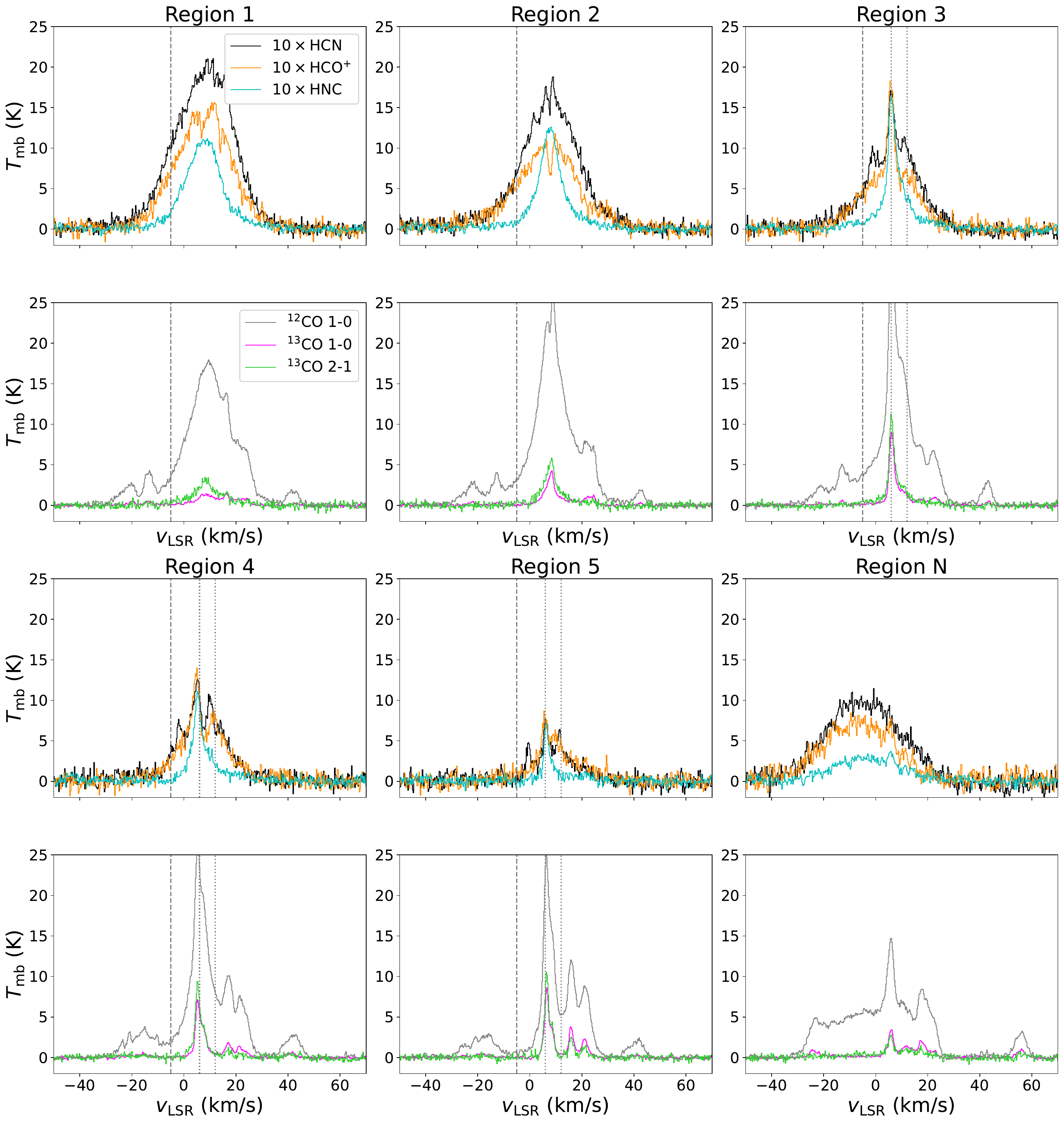}
\caption{Spectra of \hcop\ $J=1\text{--}0$ (orange), \hcn\ $J=1\text{--}0$ (black), \hnc\ $J=1\text{--}0$ (cyan), \coa\ $J=1\text{--}0$ (grey), \cob\ $J=1\text{--}0$ (magenta) and \cob\ $J=2\text{--}1$ (green) lines averaged in the regions marked in Figure \ref{fig:w28_ne_hcop}. Each region takes up two vertical figures to fully display the spectra. The main beam temperature $T_{\rm mb}$ of \hcop, \hcn\ and \hnc\ is multiplied by a factor of 10 for better display. The dashed vertical lines in the figures of Regions 1 to 5 mark a velocity of $-3$ \kms, while the dotted vertical lines in the figures of Regions 3 to 5 mark a velocity of $+6$ and $+12$ \kms\ (see the text for details).  
\label{fig:w28_ne_spec}}
\end{figure*}

To further investigate the properties of the shocked molecular clouds, we extract the spectra of the regions marked in Figure \ref{fig:w28_ne_hcop}, namely Regions 1 to 5 and Region N. 
The Regions 1 to 5 are located towards the outside of the SNR to show the spatial variation of the shock interaction. 
The spectra are shown in Figure \ref{fig:w28_ne_spec}. 
Significant line broadening can be seen in the \hcop\ and \hcn\ line profiles of Regions 1 to 4 and in Region N, which originate from the disturbance of the SNR blast wave. 
The linewidth of \hcop\ decreases from Region 1 to Region 5, indicating the disturbance effect of the shock is weakened towards the outer part of the SNR. 
In Regions 1 and 2, an absorption feature is in the center of the \hcop\ spectrum, where lines of \hnc\ and \cob\ show pure emission features. 
This is caused by the cold and dense molecular gas in the line of sight \citep{Reach_Shocked_2005}. 
In Regions 3 and 4, the \hcop\ spectra exhibit double-peak features at $\sim +6$ and $\sim +12$ \kms\ (see the dotted vertical lines in Figure \ref{fig:w28_ne_spec}), with their blue sides extending to $\leq -3$ \kms\ (see the dashed vertical lines in Figure \ref{fig:w28_ne_spec}). 

\par

The spectra of \hcn\ are generally similar to the \hcop\ spectra in the shocked regions (1, 2 and N). 
The hyperfine structures of the \hcn\ $J=1\text{--}0$ line, at $-7.1$ and $+4.9$ \kms\ offset from the strongest line in the center, can be seen in Regions 3, 4 and 5, but are covered up by strong line broadening in Regions 1, 2 and N. 

\par

Similar to \hcop\ and \hcn, \hnc\ is also a dense gas tracer, but the line profiles of \hnc\ are rather different in some of the regions. 
In Regions 1 and N, the line profiles of \hnc\ are similar to those of \hcop\ and \hcn, although they are relatively weak. 
In Region 2, \hnc\ does not show absorption as \hcop\ and \hcn\ do. 
In Regions 3 to 5, the \hnc\ emission has only one peak at $\sim +6$ \kms\ but still shows non-Gaussian line profile with a sharp peak. 

\par

Although the \coa\ lines also exhibit large linewidth, several overlapping components are present and prohibit us from identifying the shocked component. 
However, the \coa\ spectra in Regions 1 to 4 show strong emission at $-3$ \kms\ (see the dashed vertical lines in Figure \ref{fig:w28_ne_spec}), and their line profiles resemble that of \hcop\ around $-3$ \kms. 
These features of the \coa\ spectra are directly related to shock perturbation. 
In Region N, the line broadening of \coa\ is more prominent and is consistent with that of \hcop. 

\par

Both \cob\ $J=1\text{--}0$ and $J=2\text{--}1$ are found in all of the selected regions. 
Strong unshocked \cob\ is found in Region 3, which is also consistent with what we present in Figure \ref{fig:w28_ne_intensity}. 
The \cob\ molecules surrounding SNRs are commonly considered as a tracer of unshocked gas in SNRs because of their small optical depth compared with \coa, but the line profiles of \cob\ in Regions 1 and 2 are moderately broadened. 
Since \citet{Nicholas_mm_2012} have found moderately broadened $\rm C^{34}S$ line close to the regions we select, it is possible that \cob\ can show broadened line profile originated from shock perturbation, since both $\rm C^{34}S$ and \cob\ are rare isotopes and can trace dense gas. 

\subsection{ATLASGAL clump G6.796} \label{G6.796-0.256}

\begin{figure}
\centering
\includegraphics[width=0.47\textwidth]{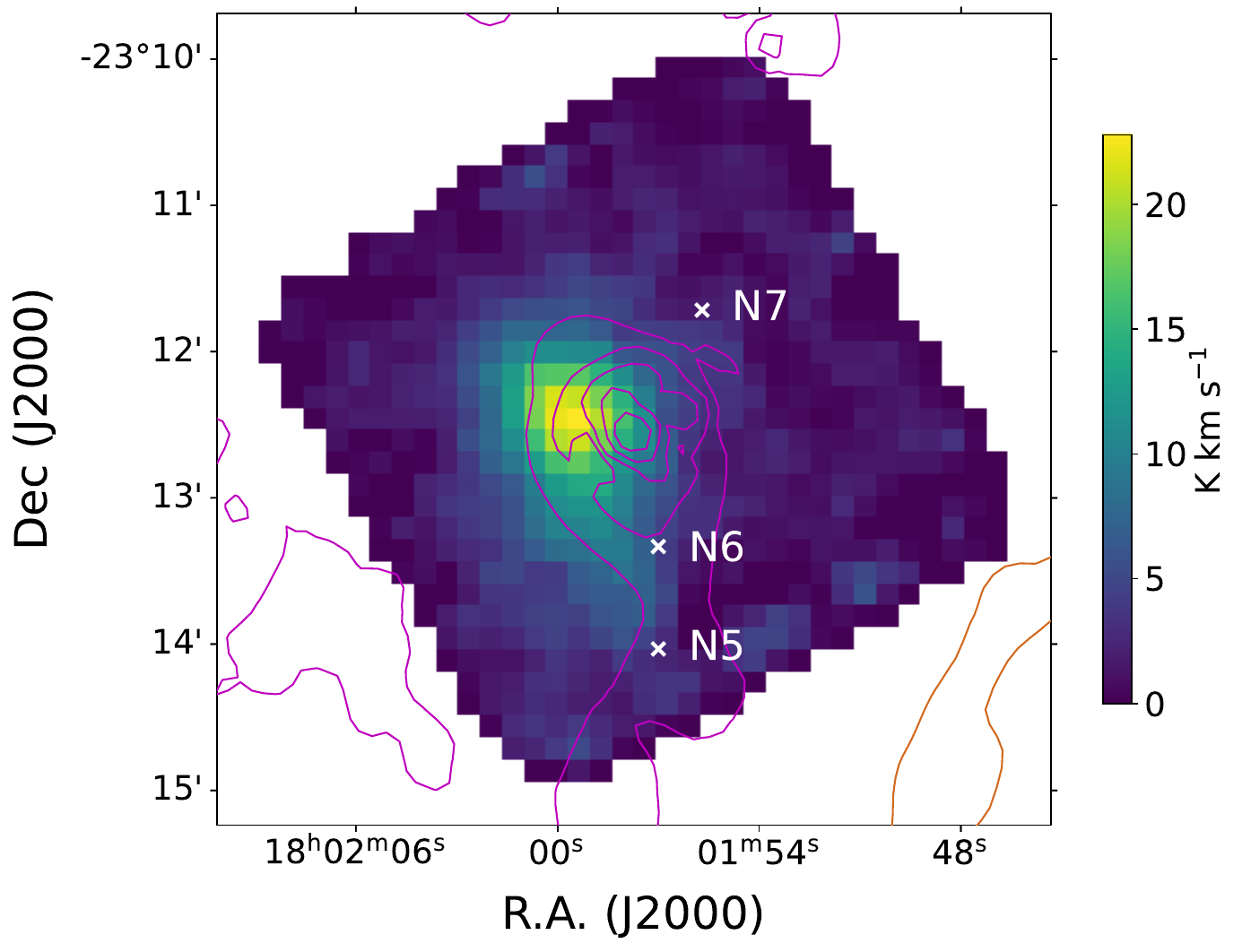}
\caption{Integrated intensity of the MALT90 \nhp\ $J=1\text{--}0$ data around clump G6.796$-$0.256 in the velocity range 0--$+35$ \kms, overlaid with magenta contours of the Herschel column density map (levels are 1.9, 2.8, 5, 10, 20 $\times 10^{22} \rm \ cm^{-2}$). The orange contours in the southwestern part of the figure is VLA 327 MHz continuum. The white crosses are the points from where \citet{Vaupre_Cosmic_2014} estimated the CR ionization rate.
\label{fig:G6.796_n2hp}}
\end{figure}

ATLASGAL clump G6.796 is a massive star-forming region \citep{Urquhart_ATLASGAL_2018} located to the northeast of the radio shell of W28 (see Figure \ref{fig:w28_ne_hcop}). 
\citet{Lefloch_Star_2008} proposed that its star-forming activity might be triggered by W28. 
The MALT90 survey has detected significant line emission of \hcop, $\rm H^{13}CO^+$, \nhp, \hcn, \hnc, $\rm C_2H$ and $\rm HC_3N$ from this clump. 
We plot the integrated intensity map of \nhp\ line emission overlaid with contours of the Herschel column density map in Figure \ref{fig:G6.796_n2hp}. 
The clump is highly compact and barely resolved by the beam of Mopra ($\sim 38^{\prime\prime}$). 
But it is clear that the peak of the molecular emission is offset from the peak of the dust emission represented by G6.796. 

\begin{figure*}
\plotone{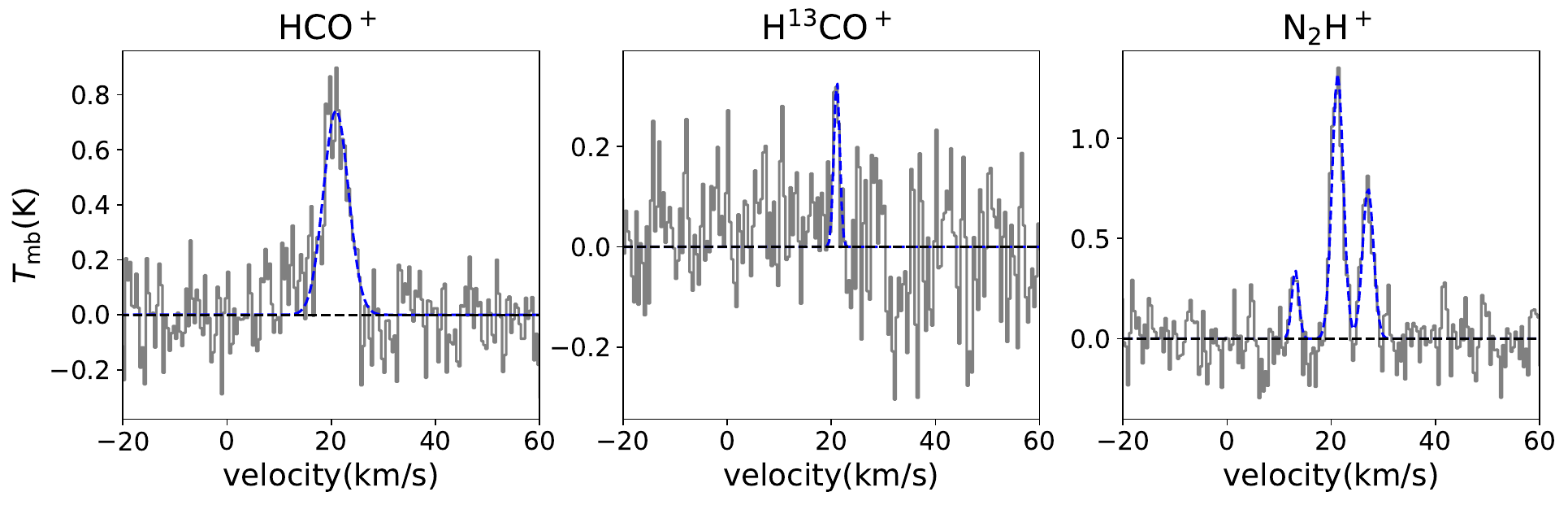}
\caption{Spectra (solid grey lines) of \hcop, $\rm H^{13}CO^+$ and \nhp\ $J=1\text{--}0$ emission averaged in a $27^{\prime\prime}\times 27^{\prime\prime}$ region towards point N6 in clump G6.796. Results of Gaussian fitting are shown with blue dashed lines. 
\label{fig:N6_spec}}
\end{figure*}

Figure \ref{fig:G6.796_n2hp} also shows the positions of three points, N5, N6 and N7, from where \citet{Vaupre_Cosmic_2014} obtained high CR ionization rate ($\gtrsim 10^{-15}\ \rm s^{-1}$). 
Among the three points, only N6 shows prominent \hcop\ and \nhp\ line emissions. 
The spectra of the \hcop, $\rm H^{13}CO^+$ and \nhp\ lines averaged in a $27^{\prime\prime}\times 27^{\prime\prime}$ region towards N6 and the results of Gaussian fitting are shown in Figure \ref{fig:N6_spec}. 
We report only a marginal detection of $\rm H^{13}CO^+$, which is inconsistent with the result of \citet{Vaupre_Cosmic_2014} who obtained a peak temperature $T_{\rm peak}$ of 0.53 K for $\rm H^{13}CO^+$. 
This may result from the different beam sizes of the MALT90 data ($38^{\prime\prime}$) and the IRAM 30m data of \citet{Vaupre_Cosmic_2014} ($29^{\prime\prime}$), as well as the relatively low sensitivity of the MALT90 data. 
The $T_{\rm peak}$ of $\rm H^{13}CO^+$ in the MALT90 data is $\approx0.33\rm \ K$. 
If the inconsistency of is due to the beam dilution effect, we expect that the $T_{\rm peak}$ detected by IRAM 30m should be approximately $0.33\times(38/29)^2 = 0.57\rm \ K$, which is roughly consistent with the result of \citet{Vaupre_Cosmic_2014}. 
The spectrum of the \nhp\ emission shows clear hyperfine structures. 

\subsection{Southern molecular clouds/HESS J1800$-$240}

\begin{figure*}
\centering
\includegraphics[width=0.7\textwidth]{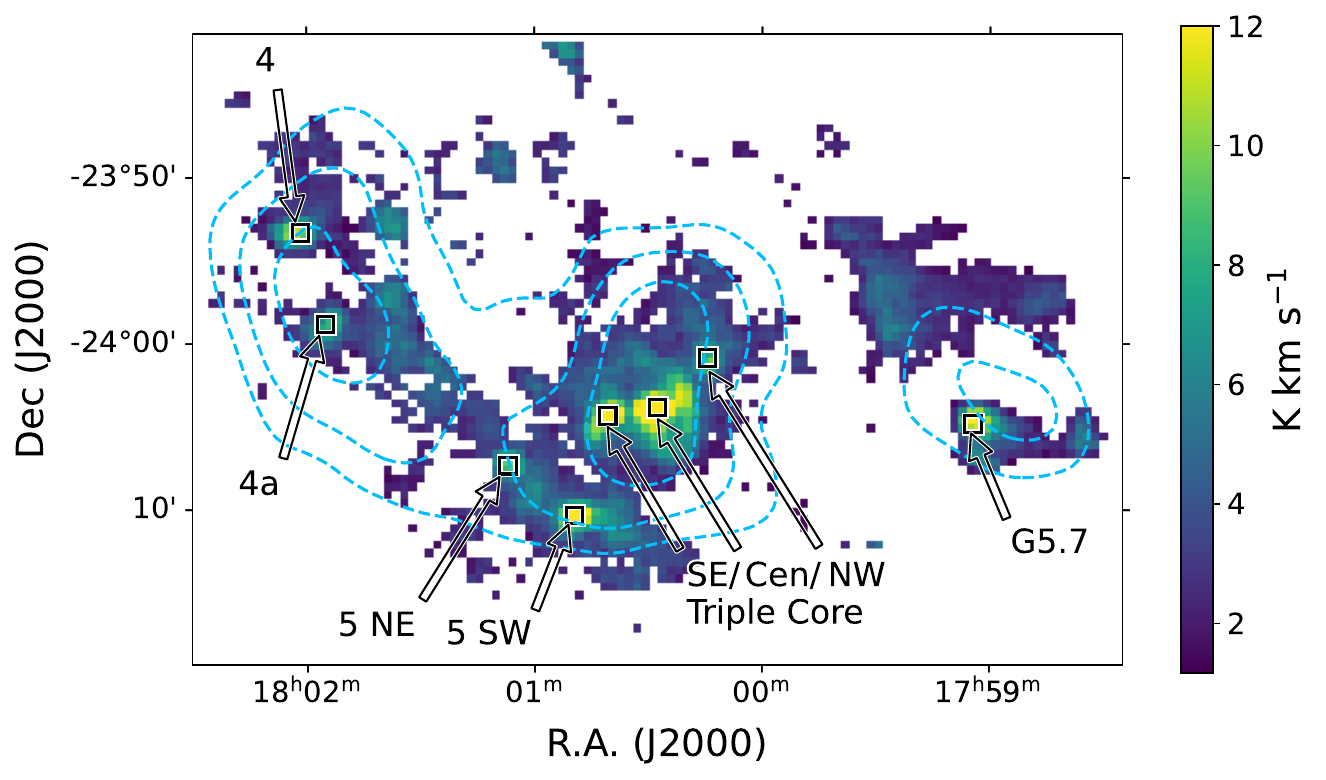}
\caption{Integrated intensity of \hcop\ $J=1\text{--}0$ from $0$ \kms\ to $+30$ \kms. The dashed blue contours trace the TeV emission of HESS J1800$-$240 (levels are 4, 5 and 6$\sigma$). The black boxes mark the regions discussed in the text.
\label{fig:w28_s_hcop}}
\end{figure*}

In Figure \ref{fig:w28_s_hcop}, we show the integrated intensity map of the \hcop\ emission from 0 \kms\ to $+30$ \kms\ in the southern molecular clouds overlaid with the HESS TeV $\gamma$-ray emission. 
This velocity interval covers all the \hcop\ emission in this region.
Clumpy structures are found widespread in the extended region and are marked with orange boxes. 
Except for clump G5.7, all other molecular clumps revealed by our \hcop\ line observation are coincident with the molecular cores found in \citet{Nicholas_12_2011}, and thus we follow their nomenclature. 
All of the marked clumps including G5.7 are related to infrared sources found by the ALTASGAL survey \citep{Urquhart_ATLASGAL_2018}. 
Cores 4 and 4a are probably related to HII regions G6.225$-$0.569 and G6.1$-$0.6, respectively \citep{Nicholas_12_2011}. 
Triple cores SE and Cen are coincident with ultra-compact HII regions G5.89$-$0.39A and B, in which G5.89$-$0.39B harbours an explosive outflow confirmed by the ALMA high-resolution observation \citep{Zapata_Confirming_2020}.
Clump G5.7 is classified as a young stellar object in \citet{Urquhart_ATLASGAL_2018} using infrared observations. 
However, it is also spatially coincident with a supernova remnant, G5.7$-$0.1 \citep{Brogan_Discovery_2006}, and an 1720 MHz OH maser \citep{Hewitt_Discovery_2009a}, which definitely shows shock-cloud interaction.

\begin{figure*}
\centering
\includegraphics[width=0.98\textwidth]{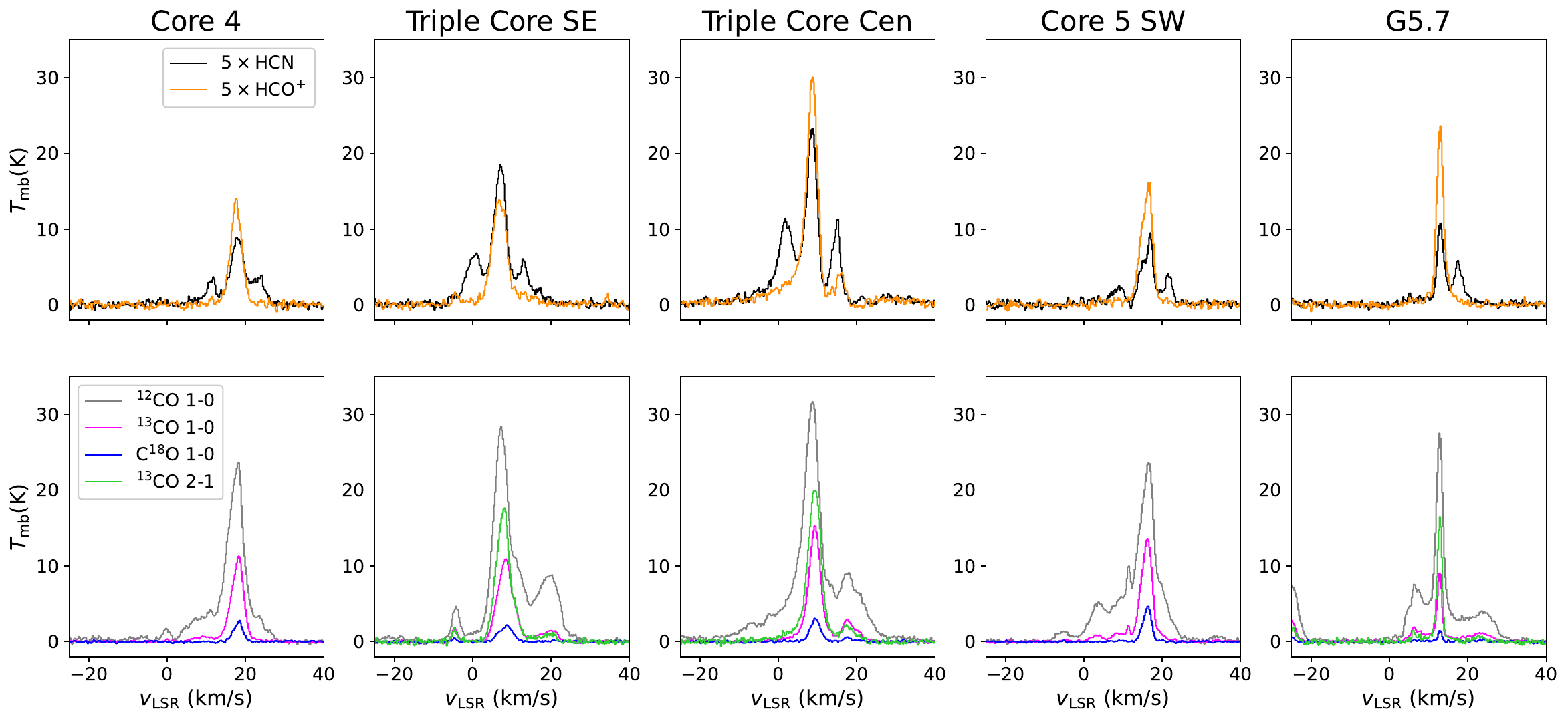}
\caption{Spectra of the \hcop\ $J=1\text{--}0$ (orange), \hcn\ $J=1\text{--}0$ (black), \coa\ $J=1\text{--}0$ (grey), \cob\ $J=1\text{--}0$ (magenta), \cob\ $J=2\text{--}1$ (green), and \coc\ $J=1\text{--}0$ (blue) lines averaged in the regions marked in Figure \ref{fig:w28_s_hcop}. The $T_{\rm mb}$ of \hcop\ and \hcn\ is multiplied by a factor of 5 for better display. In core 4, no \cob\ $J=2\text{--}1$ is shown because SEDIGISM did not cover this region. 
\label{fig:w28_s_spec}}
\end{figure*}

In Figure \ref{fig:w28_s_spec} we show the spectra of \hcop\ $J=1\text{--}0$, \hcn\ $J=1\text{--}0$, \coa\ $J=1\text{--}0$, \cob\ $J=1\text{--}0$, and \cob\ $J=2\text{--}1$ lines in core 4, triple core SE, triple core Cen, core 5 SW and clump G5.7. 
A significant contrast to the spectra in the northeastern molecular clouds is that all of the spectra in the selected regions show relatively narrow line width. 
The hyperfine structures of the \hcn\ line are clear in all of the regions. 
In the \coa\ spectra of some regions, a strong peak with $T_{\rm peak}>20$ K stands out with weaker emission in its blue and red sides. 
We tend to attribute them to irrelevant unshocked components instead of line wings because most of them show counterparts in \cob\ $J=1\text{--}0$ lines. 
In all of the regions, emission of \coc\ is found.

\par

In triple core Cen, significant asymmetric line profiles of \coa\ and \hcop\ can be seen in the blue side of the main peak. 
It is likely to be associated with the explosive outflow of G5.89$-$0.39B. 

\subsection{Line ratios}
Molecular line ratio diagnostic is often used to investigate the physical and chemical condition of the ISM. 
We herein show the line ratio maps of the line ratios $I({\rm HCN})/I({\rm HNC})$ and \cobratio\ to figure out the properties of MCs. 

\begin{figure}
\centering
\includegraphics[width=0.47\textwidth]{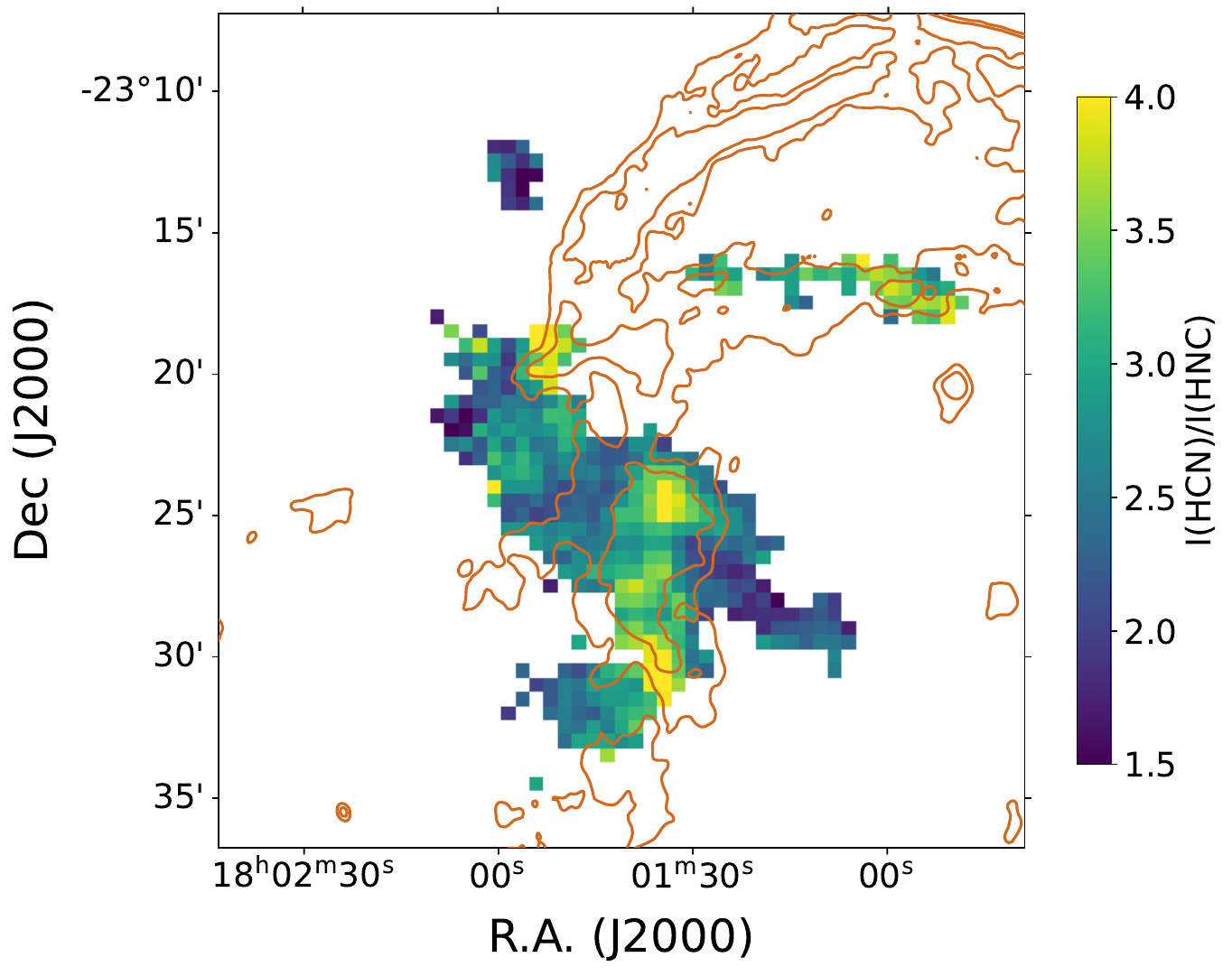}
\caption{$I({\rm HCN})/I({\rm HNC})$ ratio map of the northeastern MCs integrated in velocity range $-30$ to $+40$ \kms, overlaid with contours of the VLA 327 MHz continuum emission. 
\label{fig:w28_ne_hcnonhnc}}
\end{figure}

Line ratio $I({\rm HCN})/I({\rm HNC})$ has been proposed to be a tracer of molecular gas temperature \citep{Hacar_HCNtoHNC_2020}. 
We present the $I({\rm HCN})/I({\rm HNC})$ ratio map of the northeastern MCs in Figure \ref{fig:w28_ne_hcnonhnc}. 
The ratio $I({\rm HCN})/I({\rm HNC})$ reaches the highest ($\gtrsim 4$) towards the center of the shocked cloud. 
High value of $I({\rm HCN})/I({\rm HNC})$ is also found near Region N (defined in Figure \ref{fig:w28_ne_hcop}) and W28F. 
The spatial coincidence of high $I({\rm HCN})/I({\rm HNC})$ ratio and shocked clumps indicates shock heating. 
\citet{Hacar_HCNtoHNC_2020} found a two-part linear function between the ratio $I({\rm HCN})/I({\rm HNC})$ and the gas kinetic temperature $T_{\rm k}$ in MCs:
\begin{align}
    \begin{split}
        T_{\rm k}({\rm K}) = \left \{
        \begin{array}{ll}
            10\times \frac{I({\rm HCN})}{I({\rm HNC})} & {\rm if}\ \frac{I({\rm HCN})}{I({\rm HNC})} \leq 4 \\
            3\times \left(\frac{I({\rm HCN})}{I({\rm HNC})}-4\right) +40 & {\rm if}\ \frac{I({\rm HCN})}{I({\rm HNC})}>4
        \end{array}
        \right.
    \end{split}
\end{align}
For $I({\rm HCN})/I({\rm HNC})\sim 4$, we get $T_{\rm k}\sim 40$ K, which is roughly consistent with the temperature derived by \citet{Maxted_Ammonia_2016} with $\rm NH_3$ observation. 
These results support line ratio $I({\rm HCN})/I({\rm HNC})$ as a tracer of kinetic temperature. 

\begin{figure}
\centering
\includegraphics[width=0.47\textwidth]{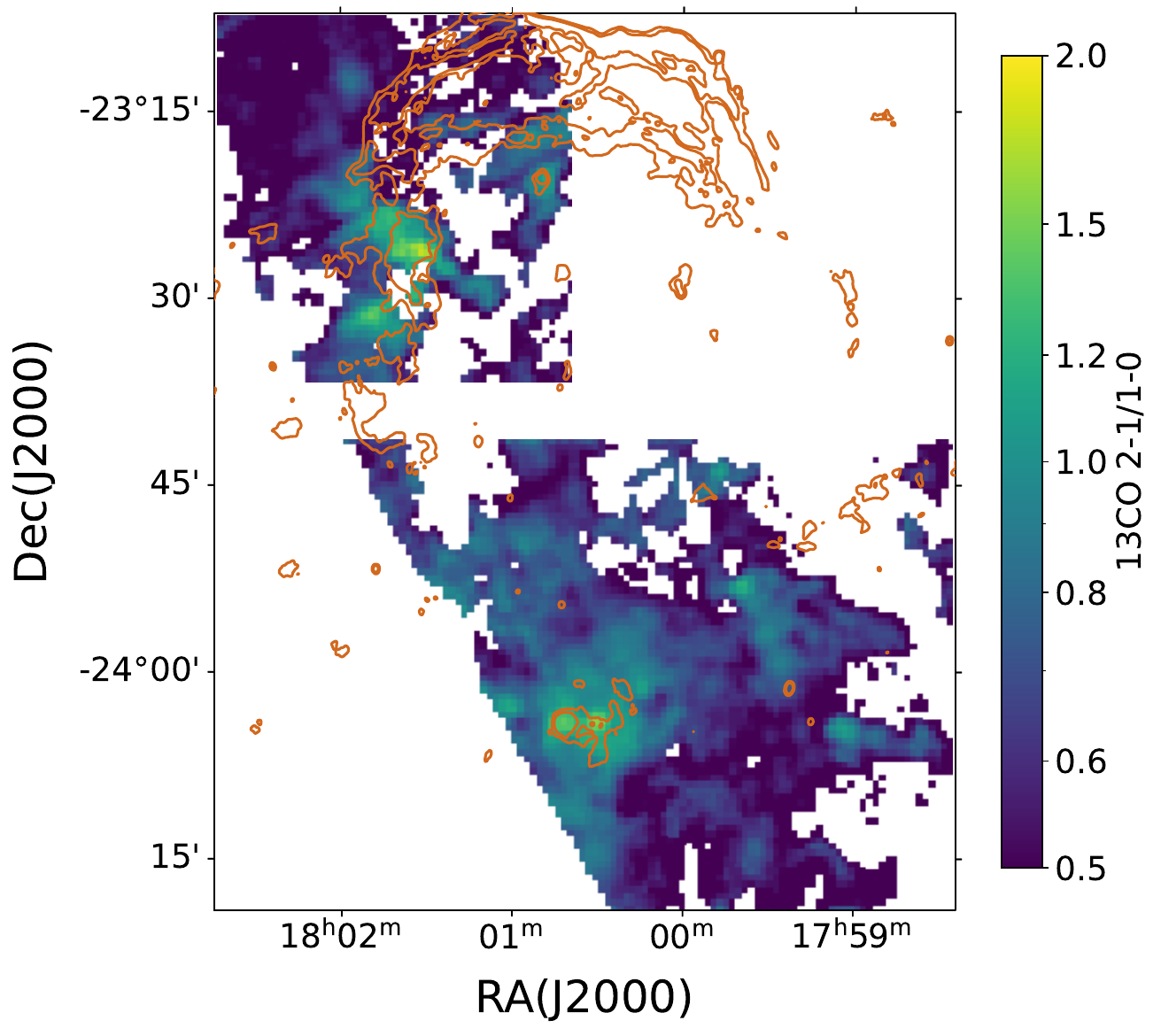}
\caption{The \cobratio\ ratio map overlaid with contours of VLA 327 MHz continuum emission. The velocity range is $-30$ to $+40$ \kms\ for the northeastern molecular clouds and $0$ to $+30$ \kms\ for the southern molecular clouds. 
\label{fig:w28_13co21on10}}
\end{figure}

Line ratio \cobratio\ is also a tracer of gas temperature since it involves two transition lines of one molecule. 
We present the \cobratio\ line ratio map in Figure \ref{fig:w28_13co21on10}. 
An enhancement of the line ratio is shown in the shocked cloud in the northeastern molecular clouds. 
However, the spatial distribution of \cobratio\ is different from that of $I({\rm HCN})/I({\rm HNC})$. 
Relatively high value is also found in the UC-HII region G5.89$-$0.39A and B, i.e. Triple core SE and Cen as labeled in Figure \ref{fig:w28_s_hcop}, which might arise from the heating of radiation and the explosive outflow. 

\par

The line ratio $I(^{12}{\rm CO}\ J=2\text{--}1)/I(^{12}{\rm CO}\ J=1\text{--}0)$ has long been used as a tracer of shock heating \citep[e.g.][]{Seta_Enhanced_1998, Jiang_Cavity_2010}. 
However, \cobratio\ can act as the same role. 
This diagnostic provides another usable way to identify SNR-cloud interaction and other heating processes, which could take the advantage of the MWISP \citep{Su_Milky_2019} and SEDISISM \citep{Schuller_SEDIGISM_2021} surveys towards the inner Galaxy. 
But we caution that $I(^{13}{\rm CO}\ J=2\text{--}1)/I(^{13}{\rm CO}\ J=1\text{--}0)$ also depends on density and other excitation effects. 
Morphological agreement between the shock front and the enhanced line ratio should be taken into consideration when using this evidence. 

\section{Discussion} \label{sec:disc}
\subsection{Column densities and Abundance ratio between \coa\ and \hcop\ } \label{sec:HCO+CO}

The abundance ratio between \hcop\ and \coa\ (hereafter CO if not specifed) is a tracer of the CR ionization rate \citep[e.g., ][]{Zhou_Unusually_2022b, Bisbas_PDFCHEM_2023}. 
To further test this notion in the SNR W28, we calculate the column density of \hcop\ and CO and their abundance ratio. 
In the shocked region of the northeastern MCs, we assume the emission of \hcop\ and CO is optically thin because of their large linewidths (which in turn mean large velocity gradients). 
We assume local thermodynamic equilibrium (LTE) in our calculation. 
For an optically thin transition, the column density of a molecular species is \citep{Mangum_How_2015}:
\begin{equation}
\begin{split}
    N_{\rm thin} = \left( \frac{3h}{8\pi^3S\mu^2R_i} \right) \left( \frac{Q_{\rm rot}}{g_Ig_Jg_K} \right) \left( \frac{\exp{(E_{\rm u}/kT_{\rm ex})}}{\exp{(h\nu/kT_{\rm ex})}-1} \right) \\
    \times \frac{1}{J_\nu(T_{\rm ex})-J_\nu(T_{\rm bg})}  \int \frac{T_{\rm R}}{f}\, dv, 
    \label{eq:thin}
\end{split}
\end{equation}
where $Q_{\rm rot}\approx kT/hb+1/3$ is the rotational partition function, $J_\nu(T)=(h\nu/k)/[\exp{(h\nu/kT)}-1]$ is the Rayleigh-Jeans equivalent temperature, and $f$, the beam filling factor, is assumed here to be unity.
Other parameters were explained in detail by \citet{Mangum_How_2015}. 
Under the LTE assumption, the excitation temperature $T_{\rm ex}$ is equal to the kinetic temperature $T_{\rm k}$. 
We adopt the kinetic temperature estimated by \citet{Nicholas_12_2011} and \citet{Maxted_Ammonia_2016} with $\rm NH_3$ observations. 
$\rm NH_3$ is a good thermometer of dense molecular gas \citep{Ho_Interstellar_1983}. The molecular constants are taken from the Cologne Database for Molecular Spectroscopy (CDMS)\footnote{\url{https://cdms.astro.uni-koeln.de/cdms/portal/}}. 

\par

To distinguish the emission of the shocked cloud from the intricate overlapping spectra, especially for the case of \coa, we carry out multi-Gaussian fitting of the spectra. 
We present the fitting results in Table \ref{tab:cden} and Appendix \ref{appen:fit}. 
Our fitting is based on the assumptions that (1) each \cob\ $J=1\text{--}0$ component should have a corresponding component of \coa, (2) the shocked \coa\ component does not have \cob\ counterpart, and (3) the spectrum of the shocked \coa\ component can be approximated as a Gaussian. 
Although assumptions (1) and (2) hold true in most cases, we note that assumption (3) is just an expedient, which may bring extra uncertainty. 

\begin{deluxetable*}{cccccccc}
\tablecaption{Results from multi-Gaussian fitting of the emission lines of \coa, \coc, \hcop\ and $\rm H^{13}CO^+$ towards two regions in the northeastern molecular clouds and six regions in the southern molecular clouds.
\label{tab:cden}}
\tablehead{ \colhead{Position} & \colhead{$T_{\rm k}$(K)$^{\rm a}$} & \colhead{Molecule} & \colhead{$v_0$(\kms)} & \colhead{$T_{\rm peak}$(K)} & \colhead{FWHM(\kms)} & \colhead{$N$($\rm cm^{-2}$)} & \colhead{\NhcoponNco$^{\rm b}$}}
\startdata 
\multicolumn{8}{c}{Northeastern molecular clouds}   \\ \hline
\multirow{4}{*}{Region 3}        & \multirow{2}{*}{55}                                                          & \coa\ (broad)              & 4.70$\pm$0.40                                                           & 5.87$\pm$0.21                                                                 & 24.74$\pm$0.53                                                         & 4.2$\times 10^{17}$                                                         & \multirow{2}{*}{1.6$\times 10^{-4}$}   \\
                                 &                                                                              & \hcop (broad)              & 5.48$\pm$0.15                                                           & 0.73$\pm$0.01                                                                 & 24.18$\pm$0.40                                                         & 6.6$\times 10^{13}$                                                         &                                         \\
                                 & \multirow{2}{*}{20}                                                          & \coc\ (narrow)            & 6.47$\pm$0.05                                                           & 1.06$\pm$0.05                                                                 & 2.42$\pm$0.12                                                          & 1.9$\times 10^{18}$                                                         & \multirow{2}{*}{2.4$\times 10^{-6}$}   \\
                                 &                                                                              & \hcop\ (narrow)           & 5.96$\pm$0.03                                                           & 1.10$\pm$0.03                                                                 & 2.42$\pm$0.08                                                          & 4.5$\times 10^{12}$                                                         &                                         \\
\multirow{2}{*}{Region N}        & \multirow{2}{*}{55}                                                          & \coa\ (broad)              & $-3.18\pm$0.24                                                          & 5.85$\pm$0.04                                                                 & 29.72$\pm$0.62                                                         & 5.0$\times 10^{17}$                                                         & \multirow{2}{*}{1.8$\times 10^{-4}$}   \\
                                 &                                                                              & \hcop\ (broad)             & $-6.41\pm$0.23                                                          & 0.73$\pm$0.01                                                                 & 32.33$\pm$0.54                                                         & 8.9$\times 10^{13}$                                                         &                                         \\ \hline
\multicolumn{8}{c}{Southern molecular clouds}                                                                                                                                                                                                                                                                                                                                                                                                                                                           \\ \hline
\multirow{2}{*}{Core 4}          & \multirow{2}{*}{20.7}                                                        & \coc                      & 18.26$\pm$0.02                                                          & 2.55$\pm$0.04                                                                 & 2.84$\pm$0.05                                                          & 5.3$\times 10^{18}$                                                         & \multirow{2}{*}{7.5$\times 10^{-6}$}   \\
                                 &                                                                              & $\rm H^{13}CO^+$          & 18.54$\pm$0.06                                                          & 0.26$\pm$0.02                                                                 & 1.70$\pm$0.14                                                          & 4.0$\times 10^{13}$                                                         &                                         \\
\multirow{2}{*}{Core 4a}         & \multirow{2}{*}{24.9}                                                        & \coc                      & 16.03$\pm$0.03                                                          & 2.39$\pm$0.04                                                                 & 2.26$\pm$0.07                                                          & 4.5$\times 10^{18}$                                                         & \multirow{2}{*}{6.7$\times 10^{-6}$}   \\
                                 &                                                                              & $\rm H^{13}CO^+$          & 15.91$\pm$0.13                                                          & 0.16$\pm$0.02                                                                 & 1.80$\pm$0.34                                                          & 3.0$\times 10^{13}$                                                         &                                         \\
\multirow{2}{*}{Triple core Cen} & \multirow{2}{*}{32.6}                                                        & \coc                      & 9.58$\pm$0.02                                                           & 2.94$\pm$0.04                                                                 & 3.10$\pm$0.05                                                          & 9.3$\times 10^{18}$                                                         & \multirow{2}{*}{1.8$\times 10^{-5}$}   \\
                                 &                                                                              & $\rm H^{13}CO^+$          & 9.59$\pm$0.04                                                           & 0.39$\pm$0.01                                                                 & 3.43$\pm$0.10                                                          & 1.7$\times 10^{14}$                                                         &                                         \\
\multirow{2}{*}{Triple core NW}  & \multirow{2}{*}{22.4}                                                        & \coc                      & 10.39$\pm$0.03                                                          & 2.04$\pm$0.05                                                                 & 2.60$\pm$0.07                                                          & 4.1$\times 10^{18}$                                                         & \multirow{2}{*}{9.7$\times 10^{-6}$}   \\
                                 &                                                                              & $\rm H^{13}CO^+$          & 10.35$\pm$0.08                                                          & 0.21$\pm$0.02                                                                 & 2.04$\pm$0.19                                                          & 4.0$\times 10^{13}$                                                         &                                         \\
\multirow{2}{*}{Core 5 SW}       & \multirow{2}{*}{17.8}                                                        & \coc                      & 16.31$\pm$0.01                                                          & 4.54$\pm$0.04                                                                 & 2.52$\pm$0.03                                                          & 7.6$\times 10^{18}$                                                         & \multirow{2}{*}{1.5$\times 10^{-5}$}   \\
                                 &                                                                              & $\rm H^{13}CO^+$          & 16.60$\pm$0.05                                                          & 0.59$\pm$0.02                                                                 & 2.33$\pm$0.11                                                          & 1.1$\times 10^{14}$                                                         &                                         \\
\multirow{2}{*}{Core 5 NE}       & \multirow{2}{*}{12.5}                                                        & \coc                      & 15.49$\pm$0.04                                                          & 1.60$\pm$0.05                                                                 & 2.33$\pm$0.11                                                          & 2.1$\times 10^{18}$                                                         & \multirow{2}{*}{1.2$\times 10^{-5}$}   \\
                                 &                                                                              & $\rm H^{13}CO^+$          & 15.28$\pm$0.08                                                          & 0.21$\pm$0.02                                                                 & 1.79$\pm$0.20                                                          & 2.5$\times 10^{13}$                                                         &                                         \\
\enddata
\tablecomments{
$^{\rm a}$ The kinetic temperature of the broad components of Region 3 is taken from \citet{Maxted_Ammonia_2016}, while the temperature of the narrow component of Region 3 is just an assumption. The temperature of Region N is set equal to that of the Region 3 broad component because of their similar physical conditions. The kinetic temperature used for the southern MCs are taken from \citet{Nicholas_12_2011}. Under the LTE assumption, $T_{\rm k}=T_{\rm ex}$ is used to derive the column density. 
$^{\rm b}$ Column density ratio between \hcop\ and \coa.
}
\end{deluxetable*}

Among the shocked regions in the northeastern MCs, we choose Regions 3 and N for typifying the fitting because both of them show clear shocked components in the \hcop\ spectra with $\rm FWHM \gtrsim 20\ km\ s^{-1}$. 
The region sample is small but representative, while it is hard to decompose the \coa\ emissions in other regions. 
We include 7--8 components to account for the \coa\ spectra, which is reasonable because all of the assumptions mentioned in the previous paragraph are satisfied. 
In Region 3, we also notice an unshocked component of \hcop\ at $+6$ \kms\ (see Figure \ref{fig:W28_NE_specfit}). We regard it as preshock gas and calculate the column density (shown in Table \ref{tab:cden}). 
Since $\rm H^{13}CO^+$ data is not available, we assume that the \hcop\ emission is optically thin despite its small linewidth, and hence the derived column density would be a lower limit. 
We use \coc\ emission to determine the column density of the unshocked CO molecules other than using \coa\ emission that is optically thick. 

\par

For the unshocked molecular clump in the southern MCs, the optically thin assumption may not hold true for the \coa\ and \hcop\ lines. 
In these cases, we calculate the column densities of their isotopic counterparts, i.e. \coc\ and $\rm H^{13}CO^+$, which are expected to be optically thin.
We assume the isotope ratios $\rm ^{12}C/^{13}C=50$ \citep{Milam_12C_2005} and $\rm ^{16}O/^{18}O=500$ \citep{Vaupre_Cosmic_2014}. 
We conduct multi-Gaussian fitting to the \coc\ and $\rm H^{13}CO^+$ spectra of the unshocked clouds. 
The results are presented in Table \ref{tab:cden} and Appendix \ref{appen:fit}. 
Among all the unshocked molecular clump in the southern MCs, we do not fit the spectra in the triple core SE and G5.7. 
In the triple core SE, the peak of the \coc\ line comprises two components but they are difficult to decompose (see Figure \ref{fig:w28_s_spec}). 
In G5.7, no $\rm H^{13}CO^+$ data is available. 
In the other regions, the components are clearly distinguishable (see Figure \ref{fig:Sspecfit}). 
The resulting column densities and abundance ratios are listed in Table \ref{tab:cden}. 
We note that it is almost impossible to determine the uncertainty brought by the LTE assumption due to some factors such as unknown gas density, so we do not give an uncertainty estimation here. 

\par

Generally, the abundance ratio \NhcoponNco\ is of the order of $10^{-5}$ or lower in the unshocked MCs, which is consistent with the value found in cold MCs ($10^{-6}\text{--}10^{-4}$) \citep[e.g.][]{Agundez_Chemistry_2013, Miettinen_MALT90_2014, Fuente_Gas_2019}. 
In the shocked clouds where the \hcop\ and \coa\ lines show large linewidths, \NhcoponNco\ of order $ 10^{-4}$. 
Although this value is not significantly higher than the normal values, it is an order of magnitude higher than that in unshocked clouds, which indicates that there is an enhancement of $\rm HCO^+/CO$ in the shocked molecular clouds of W28. 

\par

An important formation route of \hcop\ is: 
\begin{equation}
    \rm CO + H_3^+ \longrightarrow HCO^+ + H_2,
    \label{eq:hcop}
\end{equation}
in which $\rm H_3^+$ is the key product of CR ionization. 
Therefore, the abundance of \hcop\ is sensitive to the CR ionization rate. 
In comparison, the abundance of CO does not vary significantly with CR ionization rate \citep{Albertsson_Atlas_2018,Zhou_Unusually_2022b}. 
In the northeastern MC of W28, a high CR ionization rate is detected by \citet{Vaupre_Cosmic_2014}. 
There may be a link between the high CR ionization rate and the enhancement of \NhcoponNco. 
Also, since the enhancement of \NhcoponNco\ is found in shocked clouds, we expect that shock chemistry may play an important role. 

\par

To further investigate the joint chemical effect of shock wave and CRs, we apply the Paris-Durham magnetohydrodynamic (MHD) shock model \citep{Flower_influence_2003,Flower_Interpreting_2015,Godard_Models_2019} to simulate the chemical processes in MCs subjected to an MHD shock wave. 
The model is a 1D plane-parallel MHD model coupled with more than 1000 chemical reactions of over 100 species. 
This model has been used by \citet{Gusdorf_Probing_2012} to study the multi-level CO emissions in W28F. 
The entire procedure of calculation consists of two steps. 
First, a steady state condition is obtained with a fixed density to self-consistently mimic the preshock condition. 
Then, the shock condition is exerted onto the preshock gas and the code begins to calculate both the dynamic and chemical changes of the cloud. 

\par

The parameters we choose to vary are: the preshock density of hydrogen $n_{\rm H}$, the CR ionization rate per $\rm H_2$ molecule $\zeta$, and the shock velocity $V_{\rm s}$. 
We fix the magnetic parameter, $\beta=2$, which is defined as $B_0/\mu{\rm G}=\beta \left( n_{\rm H}/{\rm cm^{-3}}\right)^{0.5}$, where $B_0$ is the strength of the initial traverse magnetic field. 
We ignore the radiation field because of the high column density found by Herschel in the northeastern MCs (see Figure \ref{fig:w28_ne_intensity}). 
The the propagation time of the cloud shock is fixed at 10 kyr, which is roughly consistent with the age of W28 \citep{Velazquez_Investigation_2002}. 
When $V_{\rm s}\geq 40$ \kms, the shock wave is J-type, which would strongly dissociate the molecules on its passage, while the shock wave is C- or CJ-type when $V_{\rm s}< 40$ \kms, which is consistent with the type given by \citet{Gusdorf_Probing_2012} towards W28F. 
We therefore limit our discussions to $V_{\rm s}\leq 40$ \kms. 
The column densities of \hcop\ and CO are obtained from integrating the number densities along the shock layer.

\begin{figure}
\centering
\includegraphics[width=0.47\textwidth]{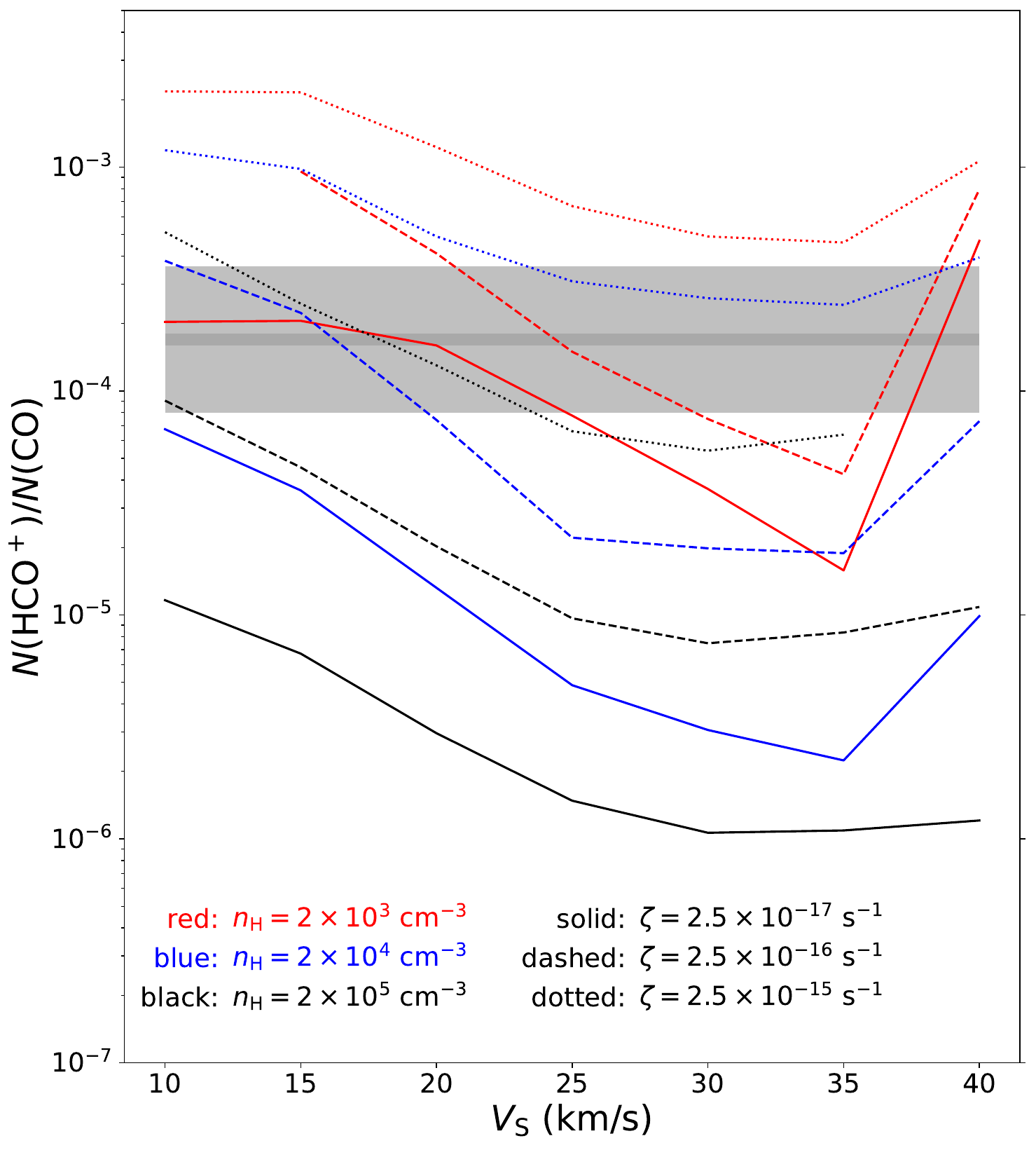}
\caption{The abundance ratio \NhcoponNco\ as a function of shock velocity, CR ionization rate, and preshock density predicted by the MHD shock code. Preshock densities of $2\times 10^3$, $2\times 10^4$, and $2\times 10^5$ $\rm cm^{-3}$ are shown in red, blue and black respectively, while CR ionization rates $2.5\times 10^{-17}$, $2.5\times 10^{-16}$, and $2.5\times 10^{-15}\rm \ s^{-1}$ are shown with solid, dashed, and dotted lines, respectively. The dark grey region indicates the observed range of \NhcoponNco: $1.6\times 10^{-4}\text{--}1.8\times 10^{-4}$ and the light grey region indicates $0.8\times 10^{-4}\text{--}3.6\times 10^{-4}$ as a rough estimation of the uncertainty. 
\label{fig:shock_results}}
\end{figure}

We present the results of the MHD simulation in Figure \ref{fig:shock_results}. 
Generally, the abundance ratio \NhcoponNco\ is higher for lower preshock density, larger CR ionization rate and smaller shock velocity. 
The model with $n_{\rm H}=2\times 10^5\rm \ cm^{-3}$, $\zeta = 2.5\times 10^{-15} \rm \ s^{-1}$ and $V_{\rm s}=15\text{--}20$ \kms\ can explain the observed abundance ratio \NhcoponNco, which is consistent with the $n_{\rm H}$ derived by \citet{Nicholas_mm_2012}, $\zeta$ derived by \citet{Vaupre_Cosmic_2014} and the shock velocity derived by \citet{Gusdorf_Probing_2012}. 
However, other combinations of parameters can also fit the observed abundance ratio \NhcoponNco\ with low CR ionization rate, but require lower density than that observed. 

\par

There are a few caveats in our comparison between the MHD simulation and the observations. 
First, the 1D simulation cannot reflect the 3D structure of the SNR. 
Second, the shock does not evolve to steady state except for high velocity shock with high density, meaning that the results of the simulation are sensitive to the propagation time of the cloud shock which is hard to determine. 
Third, the LTE assumption that the excitation temperature of both of the CO and \hcop\ emission is equal to the kinetic temperature is not necessarily satisfied in the shocked MCs, and the optically thin assumption may lead to an underestimation of the molecular column densities. 
Despite these caveats, our simulation shows that the observed $N({\rm HCO^+})$ can be explained by the parameters estimated in previous studies, and that the chemistry induced by the shock and CRs can explain the enhancement of \NhcoponNco. 

\subsection{The \NhcoponNnhp\ abundance ratio}
The abundance ratio \NhcoponNnhp\ has also been used to estimate the CR ionization rate \citep{MoralesOrtiz_Ionization_2014, Ceccarelli_Herschel_2014}. 
Both molecular species are expected to be sensitive to CR ionization rate. 
The main formation route of \nhp\ is: 
\begin{equation}
    \rm N_2 + H_3^+ \longrightarrow N_2H^+ + H_2,
\end{equation}
which is similar to that of \hcop\ (see Reaction \ref{eq:hcop}). 
The abundance ratio $N({\rm HCO^+})/N({\rm N_2H^+})$ can be used in chemical simulation to estimate the CR ionization \citep[e.g.][]{Ceccarelli_Herschel_2014}. 

\par

Enhanced CR ionization rate is detected in points N5, N6 and N7 close to the clump G6.796 (see Figure \ref{fig:G6.796_n2hp}).
These points are located outside the boundary of the W28 SNR, so they have not been disturbed by the shock wave and make it possible to investigate the pure chemical effect of CRs. 
Towards N6, we derive the column density of \hcop\ and \nhp\ using the LTE assumption and the temperature derived by \citet{Vaupre_Cosmic_2014} who used non-LTE analysis. 
Since $\rm H^{13}CO^+$ shows only marginal detection and hence the estimated $N(\rm H^{13}CO^+)$ may be unreliable, we derive two values of $N({\rm HCO^+})$: (1) $N({\rm HCO^+})=2.7\times 10^{12} \rm \ cm^{-2}$ for the optically thin case, and (2) $N({\rm HCO^+})= 50\times N({\rm H^{13}CO^+})= 1.5\times 10^{13} \rm \ cm^{-2}$ for the optically thick case. 
For \nhp, we assume that it is optically thin because the integrated intensity ratio of the blue and middle hyperfine structures (centered at 13 and 21 \kms\ respectively) is 0.16:1, which is close to the optical-thin value 0.2:1 \citep[e.g.][]{Purcell_Physical_2009}, and get $N({\rm N_2H^+})=4.5\times 10^{12} \rm \ cm^{-2}$. 
The abundance ratio \NhcoponNnhp\ is then 0.6--3.3. 
This value is within the range found in massive molecular clumps \citep{Hoq_Chemical_2013} but lower than that found in typical dark clouds like TMC 1 and L134N \citep{Agundez_Chemistry_2013, Fuente_Gas_2019}. 

\par

To further investigate the relation between the abundance ratio \NhcoponNnhp\ and the CR ionization rate, we use the public version of \textit{Nautilus}-1.1 chemical code \citep{Ruaud_Gas_2016a} to model the chemical evolution of a MC. 
\textit{Nautilus} is a three-phase gas-grain model which can compute the gas and ice abundances of molecules under various ISM conditions. 
We adopt the gas-phase reactions from kida.uva.2014 \citep{Wakelam_2014_2015}. 
The grain reactions are taken from the original \textit{Nautilus} code package. 

\begin{figure}
\centering
\includegraphics[width=0.47\textwidth]{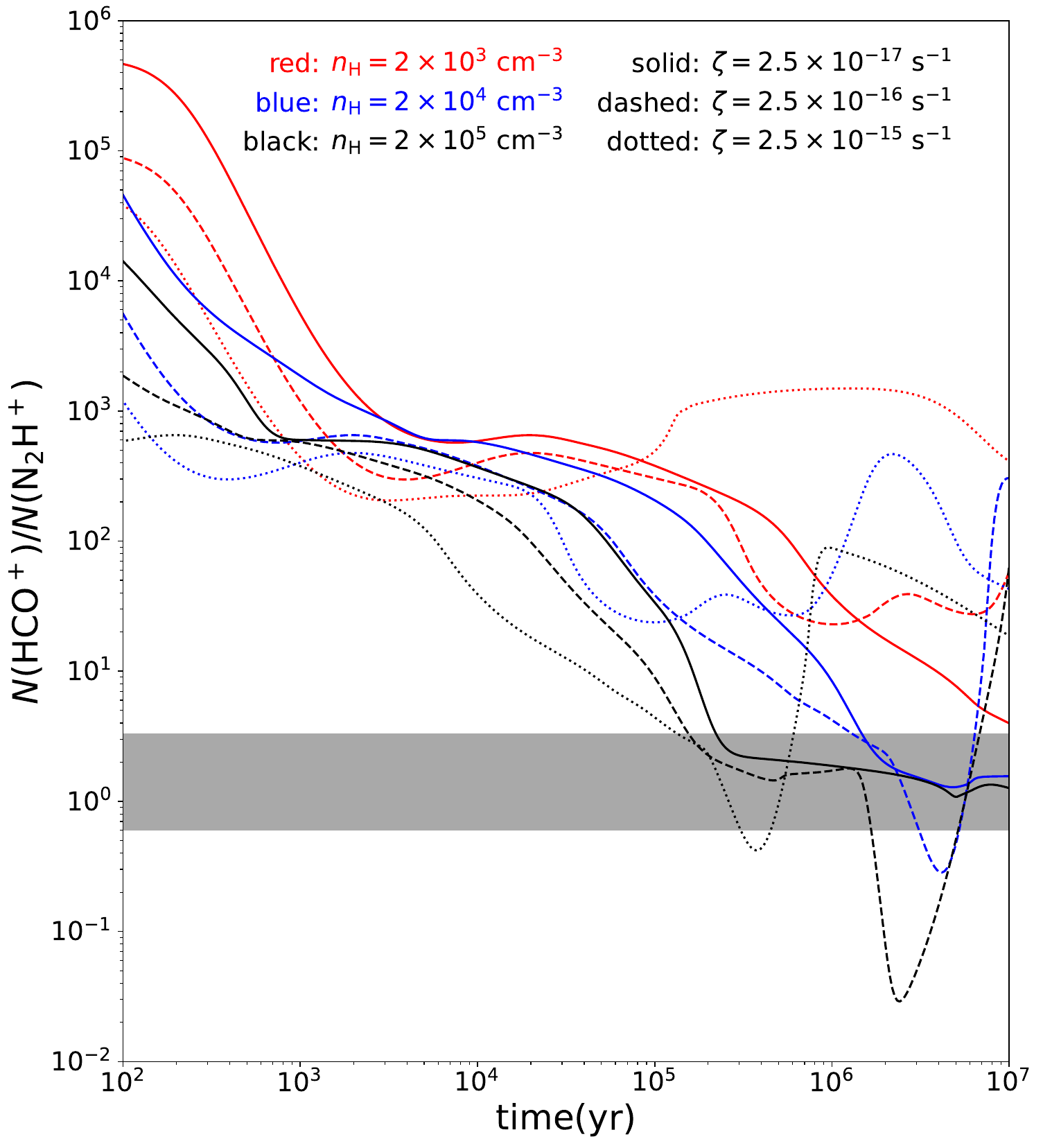}
\caption{The abundance ratio \hcop/\nhp\ as a function of density and CR ionization rate. Preshock density of $2\times 10^3$, $2\times 10^4$, and $2\times 10^5$ $\rm cm^{-3}$ are shown in red, blue and black respectively, while CR ionization rate of $2.5\times 10^{-17}$, $2.5\times 10^{-16}$, and $2.5\times 10^{-15}$ $\rm s^{-1}$ are shown with solid, dashed and dotted lines. The grey region indicates the observed range of \hcop/\nhp: 0.6--3.3.
\label{fig:nautilus}}
\end{figure}

The input parameters are the density of H nuclei ($n_{\rm H}=2\times 10^3$, $2\times 10^4$, and $2\times 10^5 \rm \ cm^{-3}$) and the CR ionization rate per $\rm H_2$ molecule ($\zeta=2.5\times 10^{-17}$, $2.5\times 10^{-16}$, and $2.5\times 10^{-15} \rm \ s^{-1}$). 
The gas temperature is fixed to 13 K which is the value obtained by \citet{Vaupre_Cosmic_2014} using non-LTE analysis towards N6. 
We set the dust temperature equal to the gas temperature. 
The evolution time is limited to $10^7$ yr. 

\par

We present the results of the chemical simulation in Figure \ref{fig:nautilus}. 
None of the modelled conditions reached a steady state. 
Four of the nine combinations of parameters can explain the observed abundance ratio \NhcoponNnhp. 
All of these parameters include high density, i.e. $n_{\rm H}\gtrsim 2\times 10^4 \rm \ cm^{-3}$. 
For $n_{\rm H} = 2\times 10^5 \rm \ cm^{-3}$, middle and high CR ionization rate is required, while for $n_{\rm H} = 2\times 10^4 \rm \ cm^{-3}$, low and middle CR ionization rate is required. 
Note that the physical conditions \citet{Vaupre_Cosmic_2014} derived for N6 is $n_{\rm H_2}=2\text{--}6 \times 10^3\rm \ cm^{-3}$ and $\zeta = 1.3\text{--}4.0 \times 10^{-15} \rm \ s^{-1}$. 
Therefore, the \NhcoponNnhp\ could not be explained by the physical parameters derived by \citet{Vaupre_Cosmic_2014}. 
However, if higher density, e.g. $n_{\rm H}=2\times 10^5 \rm \ cm^{-3}$, is adopted, the observed \NhcoponNnhp\ can be explained by high CR ionization rate $\zeta=2.5\times 10^{-15}\rm \ s^{-1}$. 

\par

Such discrepancy is expected to result from either the simulation or the observation. 
The chemical models is not always perfect because of the incomplete chemical network, uncertain reaction coefficients, undiscovered chemical processes, and unknown initial conditions, all of which are hard to deal with. 
From the observational point of view, we note that the density is derived using \cob\ $J=1\text{--}0$, $J=2\text{--}1$ and \coc\ $J=1\text{--}0$ and $J=2\text{--}1$ lines \citep{Vaupre_Cosmic_2014}. 
Using the coefficients taken from Leiden Atomic and Molecular Database\footnote{\url{https://home.strw.leidenuniv.nl/~moldata/}} \citep{vanderTak_Leiden_2020}, we find that the critical density of the used \cob\ and \coc\ lines are below $\sim 10^4 \rm \ cm^{-3}$, while the critical density of \hcop\ and \nhp\ 1--0 are both $\sim 10^5 \ \rm cm^{-3}$. 
This means that \hcop\ and \nhp\ trace the gas with higher density than the gas traced by CO isotopes, which will result in the inconsistency of the observed gas density and the density in the simulation. 
Future observations of higher-J CO transitions, whose critical densities are higher, may shed light on this problem. 

\subsection{Implications on the origin of HESS J1800$-$240}
The origin of the $\gamma$-ray emission towards HESS J1800$-$240 is still under debate. 
Some argue that all the three sources (A, B, and C) are originated from the escaped CRs from the W28 SNR \citep{Li_g-rays_2010,Cui_Leaked_2018}, while others emphasize contribution of local CR accelerators including the massive stellar cluster toward region B \citep{Gusdorf_Irradiated_2015,Hampton_Chandra_2016a} and SNR candidate G5.7$-$0.1 towards region C \citep{Joubert_FermiLAT_2016}.
If local CR contributors do exist, they are expected to drive fast shock waves that can accelerate particles to $> \text{TeV}$ energies.  
From the view of MC, we do not find either line broadening induced by shock disturbance (see Figure \ref{fig:w28_ne_spec}) nor significant shock heating (see Figure \ref{fig:w28_13co21on10}) except towards UC-HII region G5.89$-$0.39. 
However, we cannot rule out the possibility that the local accelerator is too young to have exerted significant and observable influence (line broadening and heating effect) on the MCs \citep{Sano_interstellar_2021}. 
Future high-resolution and high-sensitivity observations of multiple molecular transitions are needed to search for local CR-accelerating shock waves interacting with MCs. 

\subsection{X-ray emission towards W28F}

\begin{figure}
\centering
\includegraphics[width=0.47\textwidth]{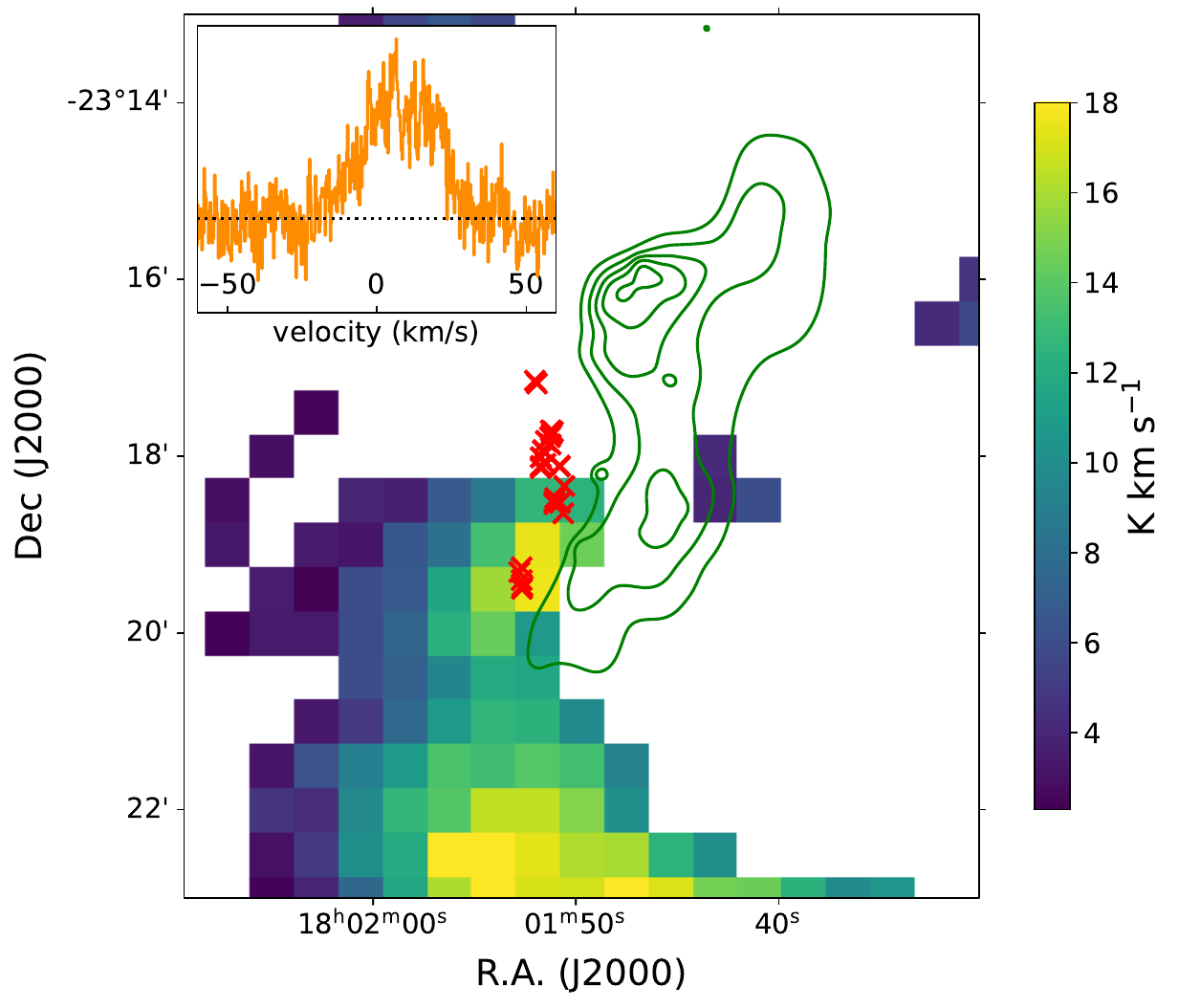}
\caption{A zoom-in view of the integrated intensity of \hcop\ from $-30$ \kms\ to $+40$ \kms\ overlaid with green contours of \textit{XMM-Newton} 0.3--7.0 keV X-ray flux map taken from \citet{Zhou_XMM-Newton_2014a}. The red crosses are 1720 MHz OH masers detected by \citet{Claussen_Polarization_1997}. An example of the \hcop\ spectrum towards the maser points is shown in the upper left subfigure. 
\label{fig:XMM}}
\end{figure}

The morphology of X-ray emission associated with W28 features an ear-like structure towards the northeastern MC \citep{Rho_ROSAT_2002}.  
In Figure \ref{fig:XMM}, we display a zoom-in view of the \hcop\ emission towards W28F overlaid with green contours of the X-ray ear. 
The X-ray emission is located on the western side of the \hcop\ emission. 
The close spatial coincidence between the molecular and X-ray emission suggests that the SNR blast wave which has heated the gas and generated an X-ray emitting plasma may be propagating into a dense molecular clump, inducing the broad \hcop\ line (see the subfigure in Figure \ref{fig:XMM}) and the 1720 MHz OH masers. 
Assuming a crude pressure balance between the cloud shock and the X-ray emitting gas, we have \citep{McKee_interaction_1975} $1.4n_m m_{\rm H} v_m^2 \sim 2.3n_{\rm H} kT$, where $n_m$ is the number density of the hydrogen atoms ahead of the cloud shock, $n_{\rm H}$ and $T$ are the hydrogen density and temperature of the X-ray emitting gas respectively ($n_{\rm H}\approx 2.7 \ \rm cm^{-3}$ and $kT\approx 0.3\rm\ keV$ according to \citet{Zhou_XMM-Newton_2014a} where $k$ is the Boltzmann constant), and $v_m$ is the velocity of the cloud shock ($\approx 20$ \kms\ according to our simulation in Section \ref{sec:HCO+CO} and \citet{Gusdorf_Probing_2012}). 
Then we obtain $n_m\sim 340 \left(n_{\rm H}/2.7\ {\rm cm^{-3}}\right) \left( kT/0.3\ {\rm keV} \right)  \left( v_m/20\ {\rm km\ s^{-1}} \right)^{-2} \rm cm^{-3}$. 
This typical density is lower than the proposed preshock density ($\sim 10^4\rm \ cm^{-3}$) given by \citet{Gusdorf_Probing_2012} and the critical density of the \hcop\ 1--0 line ($\sim 10^5\rm \ cm^{-3}$). 

\par

The production of 1720 MHz OH masers in SNR-MC interaction requires high CR ionization rate ($\sim 10^{-15}\rm \ s^{-1}$ \citep{Nesterenok_Modelling_2022}). 
However, X-ray is also possible to induce enhanced ionization rate in MCs, which facilitates the formation of OH masers \citep{Wardle_Enhanced_1999}. 
The luminosity of the X-ray ear is $L \approx 3.6\times 10^{34}\rm \ erg\ s^{-1}$ \citep{Zhou_XMM-Newton_2014a}. 
The X-ray ionization rate in the molecular clump adjacent to the X-ray ear is approximately \citep{Maloney_X-Ray_1996}:
\begin{equation} \label{eq:Xray}
    \begin{split}
        \zeta_{\rm X} \sim & 1.4\times 10^{-11}  \left( \frac{L}{10^{44} {\rm\  erg \ s^{-1}}}\right) \left( \frac{100\rm \ pc}{r}\right)^2 \\
        &\times \left( \frac{10^{22} \rm \ cm^{-2} }{N}\right) \rm \ s^{-1}.
    \end{split}
\end{equation}
Assuming the angular distance between the X-ray ear and the OH masers is $1\times$ the beam size of PMOD, i.e. $1^\prime$, and the attenuating hydrogen column density is $\sim 10^{22} \rm \ cm^{-2}$ according to the Herschel column density map (see Figure \ref{fig:w28_ne_intensity}), we obtain an X-ray ionization rate of $\sim 10^{-16}\rm \ s^{-1} $, which is lower than the CR ionization rate by an order of magnitude. 
We note that this is just a rough estimation in order of magnitude, because $\rm \zeta_{\rm X}$ depends strongly on the X-ray spectrum \citep{Zhou_Molecular_2018}, while Equation \ref{eq:Xray} applies to AGNs with harder spectra than SNRs. 
But indeed, our estimation shows that X-ray ionization does not take a dominant role in the formation of the 1720 MHz OH masers in W28F.

\section{Conclusion} \label{sec:con}
In this paper, we conduct \hcop, \hcn, and \hnc\ $J=1\text{--}0$ observations towards the MCs related to SNR W28 using the PMOD 13.7m telescope. 
With a combination of the archival data of MWISP, SEDIGISM and MALT90, we investigate the spatial distribution of the molecule emissions, their spectra, line ratios, and the chemical processes concerning their abundance ratios.
Our main findings are summarized below:

\par

1. In the northeastern MCs, strong emissions of \hcop, \hcn\ and \hnc\ are found spatially coincident with the radio continuum. 
We find broadened molecular lines with $\rm FWHM>20$ \kms, which are induced by the shock-cloud interaction. 
In the MCs south to the SNR, we find clumpy emissions of \hcop\ and \hcn. 
Spectra of Triple Core Cen (see Figures \ref{fig:w28all} and \ref{fig:w28_s_spec}) show line wing structures indicating disturbance from star formation processes. 

\par

2. Enhancement of line ratios $I({\rm HCN})/I({\rm HNC})$ and \cobratio\ in the shocked regions suggests shock heating. 
The kinetic temperature estimated with $I({\rm HCN})/I({\rm HNC})$ is roughly consistent with the temperature estimated with $\rm NH_3$ observations \citep{Maxted_Ammonia_2016}, supporting it as a thermometer of molecular gas. 

\par

3. We find that the abundance ratio \NhcoponNco\ is typically $\sim 10^{-4}$ in the shocked clouds and $\sim 10^{-5}$ in unshocked clouds. 
The Paris-Durham MHD shock code is applied to investigate the chemical effects of shock and CR ionization. 
We find that the abundance ratio \NhcoponNco\ can be reproduced with the physical parameters previously obtained from observations: preshock density $n_{\rm H}=2\times 10^5\rm \ cm^{-3}$, CR ionization rate $\zeta=2.5\times 10^{-15}\rm \ s^{-1}$ and shock velocity $V_{\rm s}=15\text{--}20$ \kms. 
Therefore, the enhancement of \NhcoponNco\ is a joint effect of shock and CR chemistry. 

\par

4. Towards point N6 (see Figure \ref{fig:G6.796_n2hp}) with known high CR ionization rate outside the radio boundary of the remnant, we estimate the abundance ratio $N({\rm HCO^+})/N({\rm N_2H^+}) \approx 0.6\text{--}3.3$. 
We use the public version of Nautilus chemical code to investigate the relation between \NhcoponNnhp\ and CR ionization rate. 
The results of the simulation show that the observed value of \NhcoponNnhp\ can be explained if higher density ($n_{\rm H}\sim 2\times 10^5 \ \rm cm^{-3}$) is adopted compared to the density derived with multiple transitions of CO isotopes, probably because \hcop\ and \nhp\ trace denser gas with lower CR ionization rate compared with CO. 
    
\par

Further high-sensitivity and high-resolution observations of multiple molecular species towards regions which exhibit shock perturbation and/or CR bombardment can help us reveal the chemical effects of shocks and CRs, and test the usability and accuracy of current chemical codes.

\begin{acknowledgments}
The authors thank the staff of Qinghai Radio Observing Station at Delingha for their help during the observation. 
T.-Y. T. thanks Qian-Cheng Liu for instructions on data reduction and analysis, Gao-Yuan Zhang and Benjamin Godard for helpful discussions, and Gavin Rowell for the HESS image of W28. Y.C. acknowledges the support from NSFC grants Nos. 12173018 and 12121003. P.Z. acknowledges the support from NSFC grant No. 12273010. S. S.-H. acknowleges support from NSERC through the Canada
Research Chairs \& the Discovery Grants program. 

\par

This research made use of the data from the Milky Way Imaging Scroll Painting (MWISP) project, which is a multi-line survey in 12CO/13CO/C18O along the northern galactic plane with PMO-13.7m telescope. We are grateful to all the members of the MWISP working group for their support. This research also made use of Montage, funded by the National Science Foundation under Grant Number ACI-1440620, previously funded by the National Aeronautics and Space Administration's Earth Science Technology Office, Computation Technologies Project, under Cooperative Agreement Number NCC5-626 between NASA and the California Institute of Technology. The SEDIGISM data products are available from the SEDIGISM survey database located at \url{https://sedigism.mpifr-bonn.mpg.de/index.html}, which was constructed by James Urquhart and hosted by the Max Planck Institute for Radio Astronomy. The data was acquired with the Atacama Pathfinder Experiment (APEX) under programmes 092.F-9315 and 193.C-0584. APEX is a collaboration among the Max-Planck-Institut fur Radioastronomie, the European Southern Observatory, and the Onsala Space Observatory.
\end{acknowledgments}

%

\vspace{5mm}

\facilities{PMO:DLH, Mopra, APEX, VLA, HESS}


\software{astropy \citep{AstropyCollaboration_Astropy_2018, AstropyCollaboration_Astropy_2022},  
          Spectral-cube \citep{Ginsburg_Radio_2015}, 
          GILDAS (Gildas Team, \url{https://www.iram.fr/IRAMFR/GILDAS/}), 
          Montage (\url{http://montage.ipac.caltech.edu/}, 
          Matplotlib (\url{https://matplotlib.org}))
          }



\appendix

\section{Results of spectral fitting} \label{appen:fit}

\begin{figure*}
\centering
\includegraphics[width=0.45\textwidth]{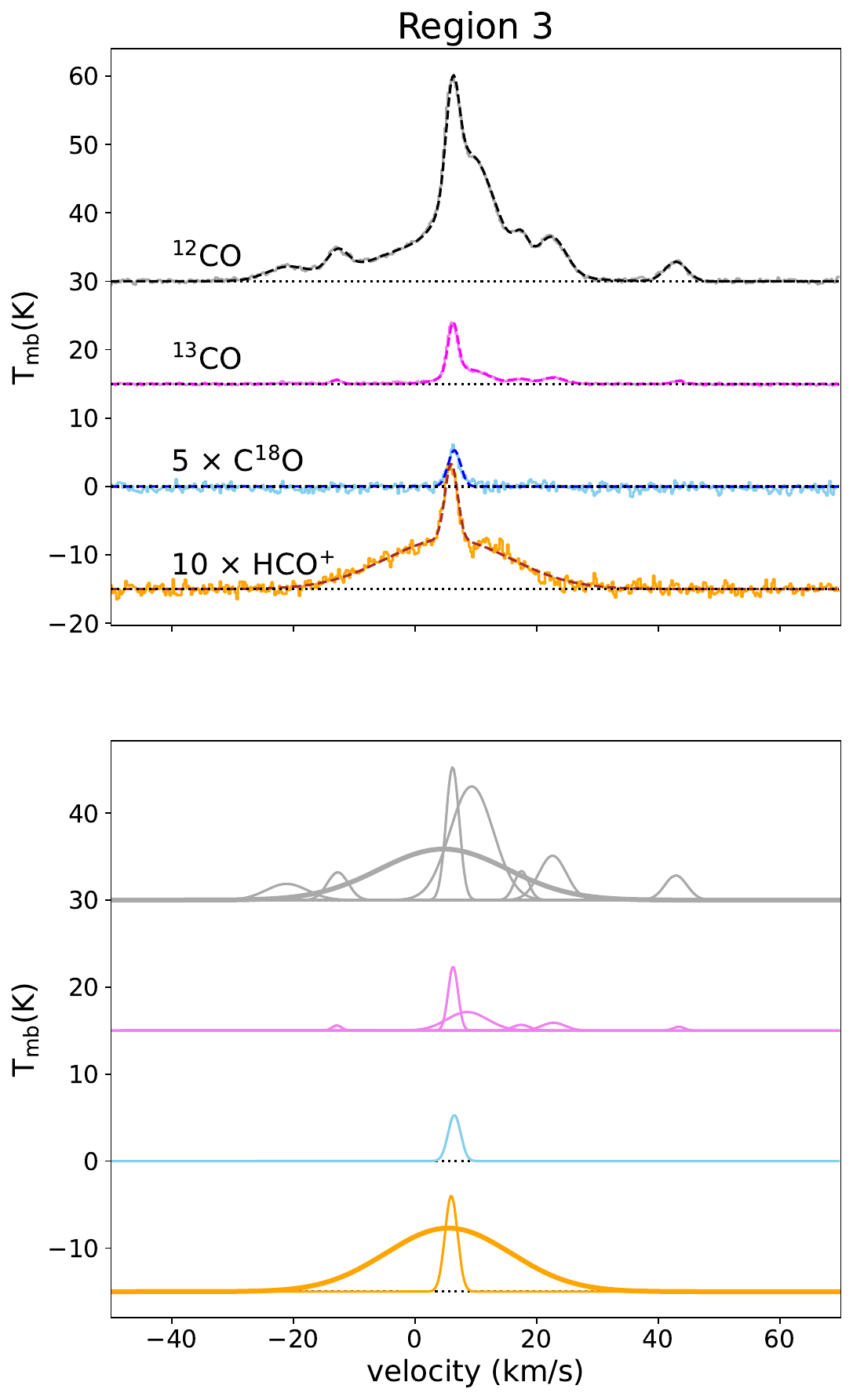}
\includegraphics[width=0.45\textwidth]{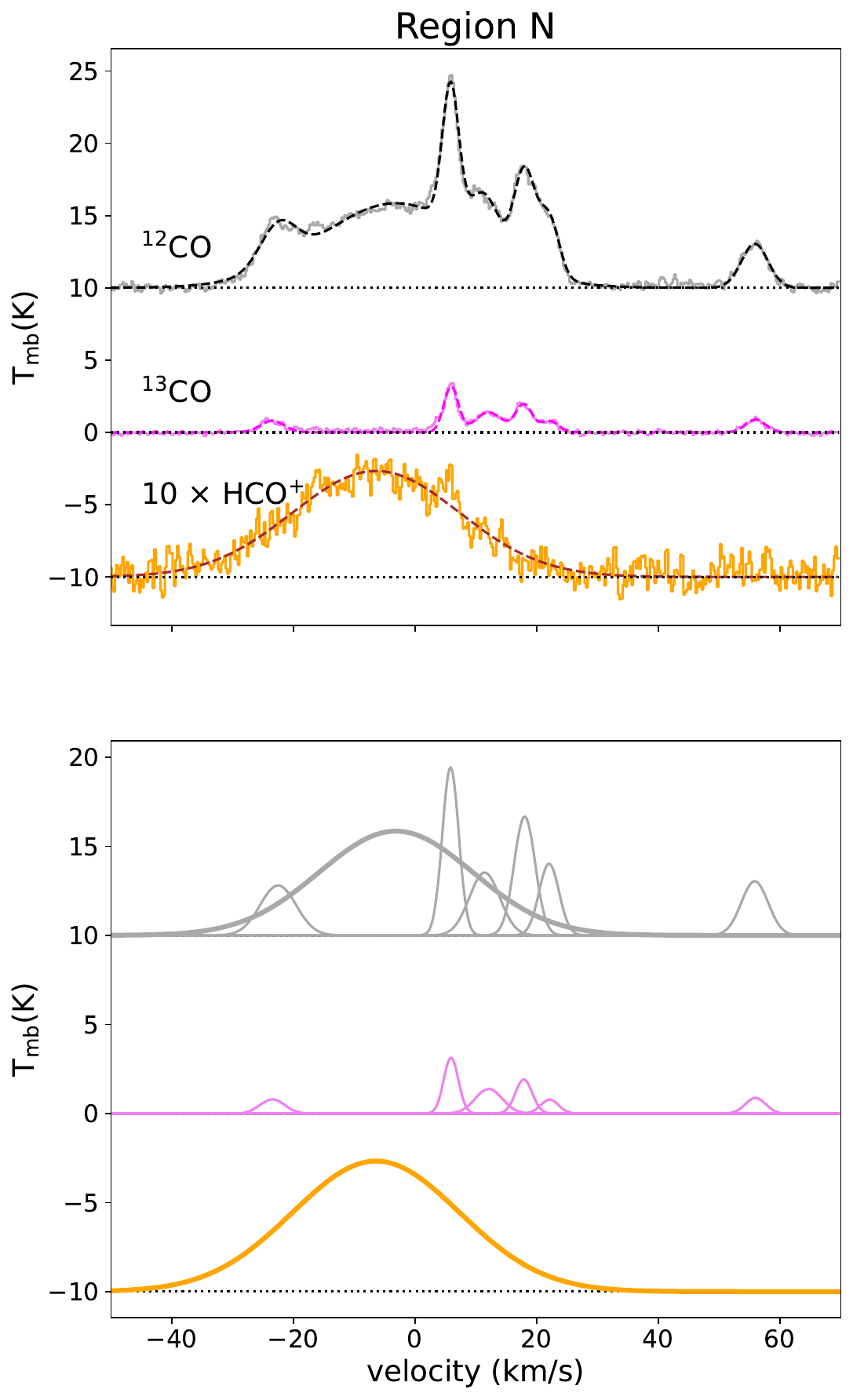}
\caption{Result of the spectral fitting for Region 3 (left panel) and Region N (right panel) in the northeastern molecular clouds of W28. In the upper panel, we show the observed spectra of \coa, \cob, \coc\ and \hcop\ 1--0 lines in grey, violet, skyblue and orange respectively. The spectra of \coc\ and \hcop\ are multiplied by a factor of 5 for better comparison. The result of multi-Gaussian fitting is presented in darker colors and dashed lines for each molecules. In the lower panel, we show the Gaussian components of each spectrum. The thicker lines of \coa\ and \hcop\ the broad lines induced by the SNR shock wave. 
\label{fig:W28_NE_specfit}}
\end{figure*}

\begin{figure*}
\centering
\includegraphics[width=0.98\textwidth]{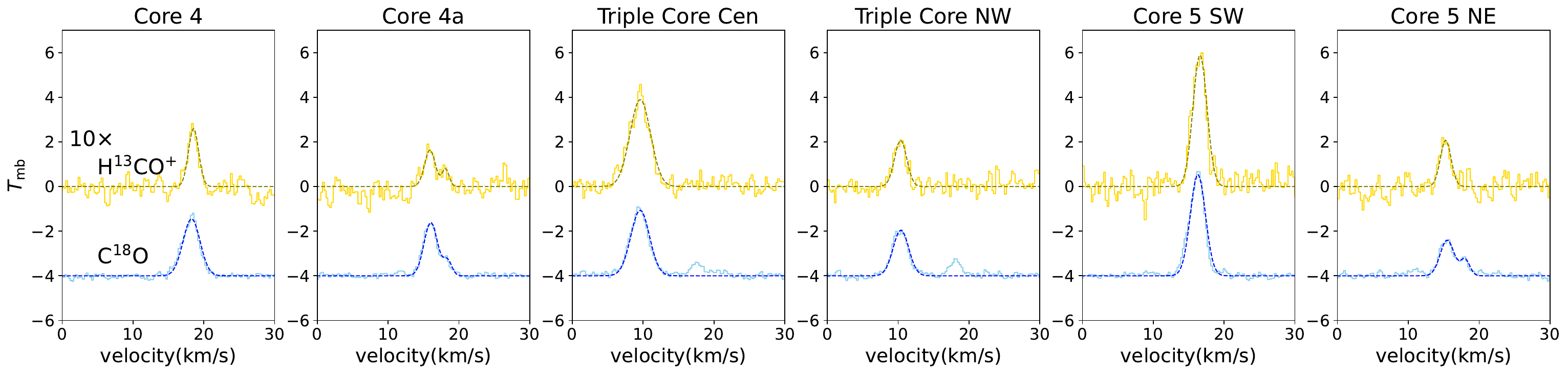}
\caption{Results of the spectra fitting of \coc\ and $\rm H^{13}CO^+$ 1--0 lines in core 4, core 4a, triple core Cen, triple core NW, core 5 SW and core 5 NE in the southern molecular clouds of W28. The observed spectra are shown in yellow and skyblue for $\rm H^{13}CO^+$ and \coc\ respectively. The results are shown with darker color and dashed lines. 
\label{fig:Sspecfit}}
\end{figure*}

In Figure \ref{fig:W28_NE_specfit}, we show the results of the spectral fitting of Region 3 and Region N in the northeastern MCs of W28. The observed results and the fitted spectra are shown in the upper panels, while the Gaussian components of each spectrum are shown in the lower panel.

\par

Generally, we get satisfactory results. Although we use four Gaussian components to account for the unshocked emissions in $-10$ to $+30$ \kms, each of the components has a corresponding \cob\ component, indicating that our estimation of the numbers of components is reasonable. We find a shocked component in \coa\ and \hcop\ lines without \cob\ counterpart in each region. 

\par

In Figure \ref{fig:Sspecfit}, we show the results of the spectral fitting of the \coc\ and $\rm H^{13}CO^+$ 1--0 lines in core 4, core 4a, triple core Cen, triple core NW, core 5 SW and core 5 NE in the southern MCs of W28. We only fit the spectra of $\rm H^{13}CO^+$ and \coc\ because the corresponding \hcop\ and \coa\ emissions are highly optically thick. In core 4a, we find two components in both of the spectra, but we use only the stronger one to calculate column density because the $\rm H^{13}CO^+$ emission is rather weak. In core 5 NE, we find two components in the \coc\ spectrum, but only one in the $\rm H^{13}CO^+$ spectrum. 


\bibliography{W28ref}{}
\bibliographystyle{aasjournal}


 
\end{CJK*}
\end{document}